\documentclass[structabstract]{aa}  

\usepackage{graphicx}
\usepackage{txfonts}
\begin{document}
   \title{An HST/WFPC2 survey of bright young clusters in M31. I.}

   \subtitle{VdB0, a massive star cluster seen at
   t$\simeq 25$ Myr.\fnmsep\thanks{Based on observations made with the NASA/ESA 
   Hubble Space Telescope, obtained at the Space Telescope Science Institute, 
   which is operated by the Association of Universities for Research in 
   Astronomy, Inc., under NASA contract NAS 5-26555. 
   These observations are associated with program GO-10818 [P.I.: J.G. Cohen].}}

   \author{S. Perina\inst{1,2}, P. Barmby\inst{3}, M.A. Beasley\inst{4,5}, 
           M. Bellazzini\inst{1}, J.P. Brodie\inst{4}, D. Burstein\inst{6},
	   J.G. Cohen\inst{7}, L. Federici\inst{1}, F. Fusi Pecci\inst{1},
	   S. Galleti\inst{1}, P.W. Hodge\inst{8}, J.P. Huchra\inst{9},
	   M. Kissler-Patig\inst{10}, 
	   T.H. Puzia\inst{11}\fnmsep\thanks{Plaskett Fellow.}
          \and
          J. Strader\inst{9}\fnmsep\thanks{Hubble Fellow.}
          }

\institute{INAF - Osservatorio Astronomico di Bologna, via Ranzani 1,
           40127 Bologna, Italy\\
	   \and
Universit\`a di Bologna, Dipartimento di Astronomia, via Ranzani 1,
           40127 Bologna, Italy\\
           \email{sibilla.perina2@unibo.it}\\
	   \and	   
Department of Physics and Astronomy, University of Western 
	      Ontario, London, ON, Canada N6A 3K7\\
	   \and	   
UCO/Lick Observatory, University of California, Santa Cruz, 
	      CA 95064, USA\\	  
	   \and	   
	      Instituto de Astrofísica de Canarias, La Laguna 38200, 
	      Canary Islands, Spain\\ 
	   \and	   
	      Department of Physics and Astronomy, Arizona State University, 
	      Tempe, AZ \\
	   \and	   
	      Palomar Observatory, Mail Stop 105-24, California Institute of Technology, 
	      Pasadena, CA 91125\\
	      \email{jlc@astro.caltech.edu}\\
	   \and	   
	      Department of Astronomy, University of Washington, Seattle, 
	      WA 98195, USA\\
	   \and	   
	      Harvard-Smithsonian Center for Astrophysics, Cambridge, MA\\
	   \and	   
	      European Southern Observatory, Karl-Schwarzschild-Strasse 2, 85748 
	      Garching bei M\"unchen, Germany\\
	  \and    
	      Herzberg Institute of Astrophysics, 5071 West Saanich Road, 
	      Victoria, BC V9E 2E7, Canada\\
             }

   \date{Received ; accepted }

 
  \abstract
   {}
   {We introduce our imaging survey of possible young massive globular clusters
   in M31 performed with the Wide Field and Planetary Camera 2 (WFPC2) on the  Hubble
   Space Telescope (HST).  We obtained shallow (to B$\sim 25$) photometry of
   individual stars in 20 candidate clusters.  We  present here details of the
   data reduction pipeline that is being applied to all the survey data and
   describe its application to the brightest among our targets, van den Bergh~0
   (VdB0), taken as a test case.}
   {Point spread function fitting photometry of individual stars was
   obtained for all the WFPC2 images of VdB0 and the completeness of the final
   samples was estimated using an extensive set of artificial stars
   experiments. The reddening, the age and the metallicity of the cluster were
   estimated by comparing the observed color magnitude diagram (CMD) with
   theoretical isochrones.  Structural parameters were obtained from 
   model-fitting to the intensity profiles measured within circular apertures on the
   WFPC2 images.}
   {Under the most conservative 
   assumptions, the stellar mass of VdB0 is $M> 2.4\times 10^4 ~M_{\sun}$,  but
   our best estimates lie in the range $\simeq 4-9\times 10^4 ~M_{\sun}$.  The
   CMD of VdB0 is best reproduced by models having solar metallicity and age
   $\simeq 25$ Myr. Ages less than $\simeq 12$ Myr and greater than $\simeq
   60$ Myr are clearly ruled out by the available data. The cluster has a
   remarkable number of red super giants ($\ga 18$) and a CMD very similar to
   Large Magellanic Cloud clusters usually classified as young globulars such
   as NGC~1850, for example.}
   {VdB0 is significantly brighter ($\ga 1$ mag) than Galactic open clusters of
   similar age. Its present-day mass  and half-light radius ($r_h=7.4$ pc) are
   more typical of faint globular clusters than of open clusters. However,
   given its position within the disk of M31, it is expected to be destroyed by
   dynamical effects, in particular by encounters with giant molecular clouds,
   within the next $\sim 4$ Gyr.}
   

   \authorrunning{S. Perina et al.}
   \titlerunning{VdB0, a massive star cluster at $t=25$ Myr.}
   \maketitle
%

\section{Introduction}

Much of the star formation in the Milky Way is thought to have occurred within
star clusters  (Lada et al. \cite{lada91};  Carpenter et al.
\cite{carp}).   Therefore, understanding the formation and evolution of star
clusters is an important piece of the galaxy formation puzzle. Our
understanding of the star cluster systems of spiral galaxies has  largely come
from studies of the Milky Way. Star clusters in our Galaxy have traditionally
been separated into two varieties, open and globular clusters  (OCs and GCs
hereafter).  OCs are conventionally regarded as young ($<10^{10}$ Gyr),
low-mass  ($<10^4 M_{\sun}$) and metal-rich systems that reside in the
Galactic disk.  In contrast, GCs are characterized as old, massive
systems. In the Milky Way,  GCs can be broadly separated into two components: a
metal-rich disk/bulge  subpopulation, and a spatially extended, metal-poor halo
subsystem  (Kinman \cite{tom}, Zinn \cite{zinn};  see also Brodie \& Strader
\cite{brodie};  Harris \cite{h01}, for general reviews  of GCs).

However, the distinction between OCs and GCs has become increasingly blurred.
For example, some OCs are sufficiently luminous and old to be confused with
GCs  (e.g., Phelps \& Schick \cite{phelps}). Similarly, some GCs are very 
low-luminosity systems (e.g., Koposov et al. \cite{kopos}) and at least one has an
age that is consistent  with the OC age distribution (Palomar 1; Sarajedini et
al. \cite{sara}). Moreover, a third category of star cluster,  ``young massive
clusters'' (YMCs) are observed to exist in both merging  (e.g., Whitmore \&
Schweizer \cite{wischw}) and quiescent galaxies  (Larsen \& Richtler
\cite{lari}), Indeed, YMCs have been known to exist in the Large Magellanic
Cloud  for over half a century (Hodge \cite{ho61}).  These objects are
significantly more luminous than  OCs (M$_V\la-8$ up to M$_V\sim -15$), making
them promising candidate young GCs. Once thought  to be absent in the Milky
Way, recent observations suggest that their census may be quite incomplete, as
some prominent cases have been found  recently in the Galaxy as well 
(Clark et al. \cite{clark}; Figer \cite{figer}).

Thus, a picture has emerged that, rather than representing distinct entities, 
OCs, YMCs and GCs may represent regions within a continuum of cluster properties
dependent upon local galaxy conditions (Larsen \cite{lars}). The lifetime of a
star cluster is  dependent upon its mass and environment. Most low-mass star
clusters in disks are rapidly disrupted via interactions with giant molecular
clouds (Lamers \& Gieles \cite{lamers}; Gieles et al. \cite{gieles}).  
These disrupted  star clusters are
thought to be the origin of much of the present field star populations (Lada \&
Lada \cite{lala}). Surviving disk clusters may then be regarded as OCs or YMCs,
depending upon their  mass. Star clusters in the halo may survive 
longer
since they are subjected to  the more gradual dynamical processes of two-body
relaxation and evaporation.  The clusters which survive for an Hubble time --
more likely to occur away from the  disk -- are termed GCs (see also
Krienke \& Hodge \cite{krie1}). 
To date, no known {\it thin} disk GCs have been identified in the Milky Way.

After the Milky Way, M31 is the prime target for expanding our knowledge of
cluster systems in spirals. 
However, our present state of knowledge about the M31
cluster system is far from complete.  Similar to the Milky Way, M31 appears to
have at least two GC  subpopulations; a metal-rich, spatially concentrated
subpopulation of GCs  and a more metal-poor, spatially extended GC
subpopulation (Huchra et al. \cite{huchra};  Barmby et al. \cite{barm00}). 
Also, again similar to
the Milky Way GCs, the metal-rich GCs in M31 rotate and  show "bulge-like"
kinematics (Perrett et al. \cite{perrett}). However, unlike the case in  the Milky Way,
the metal-poor GCs also show significant rotation (Huchra et al. \cite{huchra}; 
Perrett et al. \cite{perrett}, Lee et al.~\cite{lee}).  
Using the Perrett et al. (\cite{perrett}) data, Morrison et al. (\cite{morri})
identified what  appeared to be a {\it thin} disk population of GCs,
constituting some  27\% of the Perrett et al. (\cite{perrett}) sample. 
Subsequently,
it has been shown that at least a subset of these objects  are in fact young
($\leq$ 1 Gyr), metal-rich star clusters rather than old ``classical''  GCs
(Beasley et al. \cite{beas}; Burstein et al. \cite{burst}; 
Fusi Pecci et al.~\cite{ffp}; Puzia et al. \cite{puzia}).

Fusi Pecci et al. (\cite{ffp}; hereafter F05) presented a comprehensive 
study of bright young disk clusters in M31, selected from the Revised
Bologna Catalogue\footnote{\tt www.bo.astro.it/M31} (RBC, Galleti et
al.~\cite{rbc}) by color [$(B-V)_0\le 0.45$] or by the strength of the
$H\beta$ line in their spectra ($H\beta\ge 3.5\AA$). While these
clusters have been noted since Vetesnik (\cite{vetes}) and have been
studied by various
authors, a systematic study was lacking. 
F05 found that these clusters, that they termed -- to add to the growing 
menagerie of star cluster species -- ``Blue Luminous Compact Clusters'' 
(BLCCs), are fairly numerous in M31 (15\% of the whole GC sample),  
they have positions and kinematics typical of {\em thin disk} objects, and
their colors and spectra strongly suggest that they have ages
(significantly) less than 2 Gyr. 

Since they are quite bright ($-6.5\la M_V\la -10.0$) and -- at least in some
cases -- morphologically similar to old GCs (see Williams \& Hodge
\cite{will}, hereafter WH01), BLCCs could be regarded as YMCs, 
that is to say, candidate  young globular clusters. In particular, F05 concluded
that if most of the BLCCs have an age $\ga 50-100$ Myr they are likely
brighter than Galactic Open Clusters (OC) of similar ages, thus they
should belong to a class of objects that is not present, in large numbers, 
in our own Galaxy. 
Unfortunately, the accuracy in the age estimates obtained from the 
integrated properties of the clusters is not sufficient to determine their 
actual nature on an individual basis, i.e., to compare their total luminosity 
with the luminosity distribution of OCs of similar age (see Bellazzini et al.
\cite{cefa} and references therein).

In addition to the question of the masses and ages of these BLCCs, 
it has become clear that the BLCC photometric and spectroscopic samples
in M31 may suffer from significant contamination. 
Cohen, Matthews \& Cameron (\cite{coh}, 
hereafter C06) presented NIRC2@KeckII Laser Guide Star Adaptive Optics (LGSAO) 
images of six candidate BLCCs. Their $K\arcmin$ very-high spatial resolution 
images revealed that in the fields of four of the candidates there was
no apparent cluster. This lead C06 to the conclusion that some/many of the
claimed BLCC may in fact be just {\em asterisms}, i.e. chance groupings
of stars in the dense disk of M31. While the use of the near infrared
$K\arcmin$ band (required by the LGSAO technique) may be 
largely insensitive 
to very young clusters that are dominated by relatively few hot
stars, which emit most of the light in the blue region of the spectrum, 
the inference is that the true number of massive young clusters of M31 may 
have been severely overestimated.

Therefore, in order to ascertain the real nature of these BLCCs we have
performed  an HST survey to image 20 BLCCs in the disk of M31 (program GO-10818,
P.I.: J. Cohen).  The key aims of the survey are:

\begin{enumerate}

\item to check if the imaged targets are real clusters or asterisms, and
to determine the fraction of contamination of BLCCs by asterisms;

\item to obtain an estimate of the age of each  cluster in order to verify
whether it is brighter than Galactic OCs of similar age. Ultimately the
survey aims to provide firm conclusions on the existence of
BLCCs (YMCs) in M31 as a distinct class of object with respect to OCs (see
Krienke \& Hodge \cite{krie1,krie2}, and references therein).

\end{enumerate}

In the present contribution we describe the data reduction and analysis
strategies that we will apply to our cluster sample to estimate their
ages and metallicities. The overall procedure is described using the
brightest among the observed clusters, VdB0, as a specific case. 
We conclude this section with a brief
presentation of the cluster VdB0, below.

The present paper is organized as follows. 
The observations and the data
reduction procedure are described in detail in Sect.~2; the principal
 assumptions that will be adopted in the whole survey are also
reported in this section. Sect.~3 is devoted to the analysis of the
surface brightness profile and of the
Color Magnitude Diagram of VdB0, including total luminosity, age and
metallicity estimates. In Sect.~4 our main results are briefly
summarized and discussed.

\begin{table*}
\caption{Positional and Photometric parameters for VdB0 from the RBC$^a$}
\label{table:1}      
\centering          
\begin{tabular}{c c c c c c c c c c c c c}     
\hline\hline       
NAME & alt NAME & RA$_{J2000}$ & Dec$_{J2000}$ &X &Y & U & B & V & R & J & H & K\\ 
\hline                    
VdB0 & B195D$^b$  & 00:40:29.3 & +40:36:14.7 & -47.2$\arcmin$ &-4.3$\arcmin$& 14.97 & 15.31 & 15.06 & 14.92& 13.77 & 13.14 & 12.99\\  
\hline                  
\end{tabular}
\begin{list}{}{}
\item[$^{\mathrm{a}}$]  X and Y are projected coordinates in the direction along
(increasing Eastward) and perpendicular to the major axis of M31 
(increasing Northward) respectively, 
in arcmin, see Galleti et al. 2004, and references therein. 
\item[$^{\mathrm{b}}$] see Sect.~2.5.
\end{list}
\end{table*}

   \begin{figure}
   \centering
   \includegraphics[width=9cm]{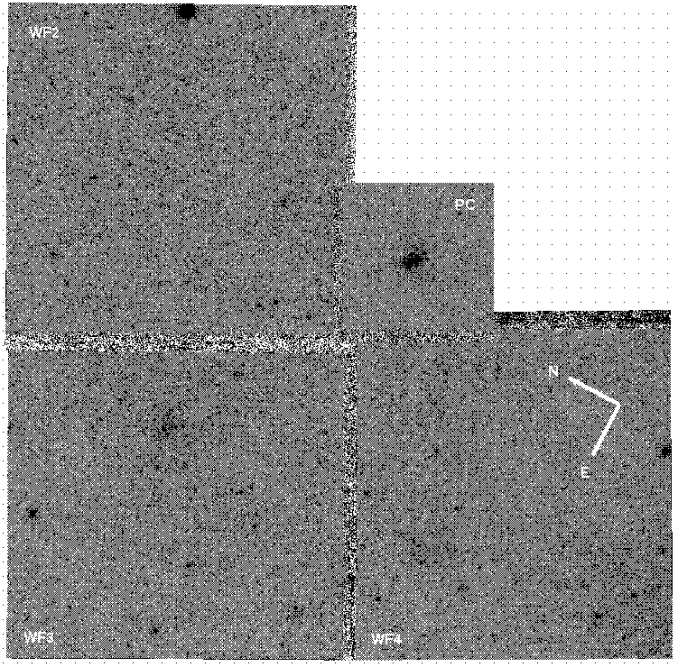}
      \caption{F450W mosaic of the whole field sampled by our WFPC2
      observations. The cluster VdB0 is at the
      center of the PC camera.}
         \label{imatut}
   \end{figure}

\subsection{The cluster van den Bergh~0 (VdB0)}

VdB0 was indicated as an open cluster by Hubble (\cite{hub36}) in the image on
the frontispiece of his book {\em The Realm of the  Nebulae}\footnote{S. van den
Bergh kindly drove our attention to this curious occurrence.}.
van den Bergh (\cite{vdb69}) presents VdB0 as the brightest open cluster of M31, 
reporting an integrated spectral type A0. He also notes that 
the cluster contains the Cepheid variable V40 (Hubble \cite{hubble}). 
A check of Hubble's (\cite{hubble}) finding charts revealed that 
two sources are labeled \# 40  in his plate VII: one of them seems indeed 
associated  with the cluster, while the other 
is $\sim 8\arcmin$  away from  VdB0, near the association  OB78 = NGC 206 
(van den Bergh \cite{vdbOB}; see also Hodge  \cite{open}). 
The cluster was re-discovered by Hodge \cite{open}, who classified 
it as an open cluster (C107, see also Hodge \cite{atlas}). Finally, 
Battistini et al. (\cite{bat87}) listed the cluster as their class D candidate 
globular cluster number 195 (B195D in the RBC). The failure to identify
B195D with VdB0 was due to the fact that 
the coordinates provided by van den Bergh (\cite{vdb69}) were in error 
by $\simeq 17\arcsec$. 
For this reason VdB0 and B195D survived as independent 
entries in M31 GC catalogues until the present day. In our 
survey we imaged both the clusters and the WFPC2 images 
revealed unequivocally that the two targets are in fact the same 
cluster. In particular the images intended to observe B195D have 
the cluster in the center of the PC camera while in the 
VdB0 images the cluster lie in the corner of the PC opposite to 
the WF cameras, such that part of the cluster is out of the image. 
In the following (and in the future) we will refer to the cluster 
as VdB0. The dataset analysed here is the one with the cluster 
centered on the PC images, hence the actual label in the header 
of the fits files is B195D. 

   \begin{figure}
   \centering
   \includegraphics[width=9cm]{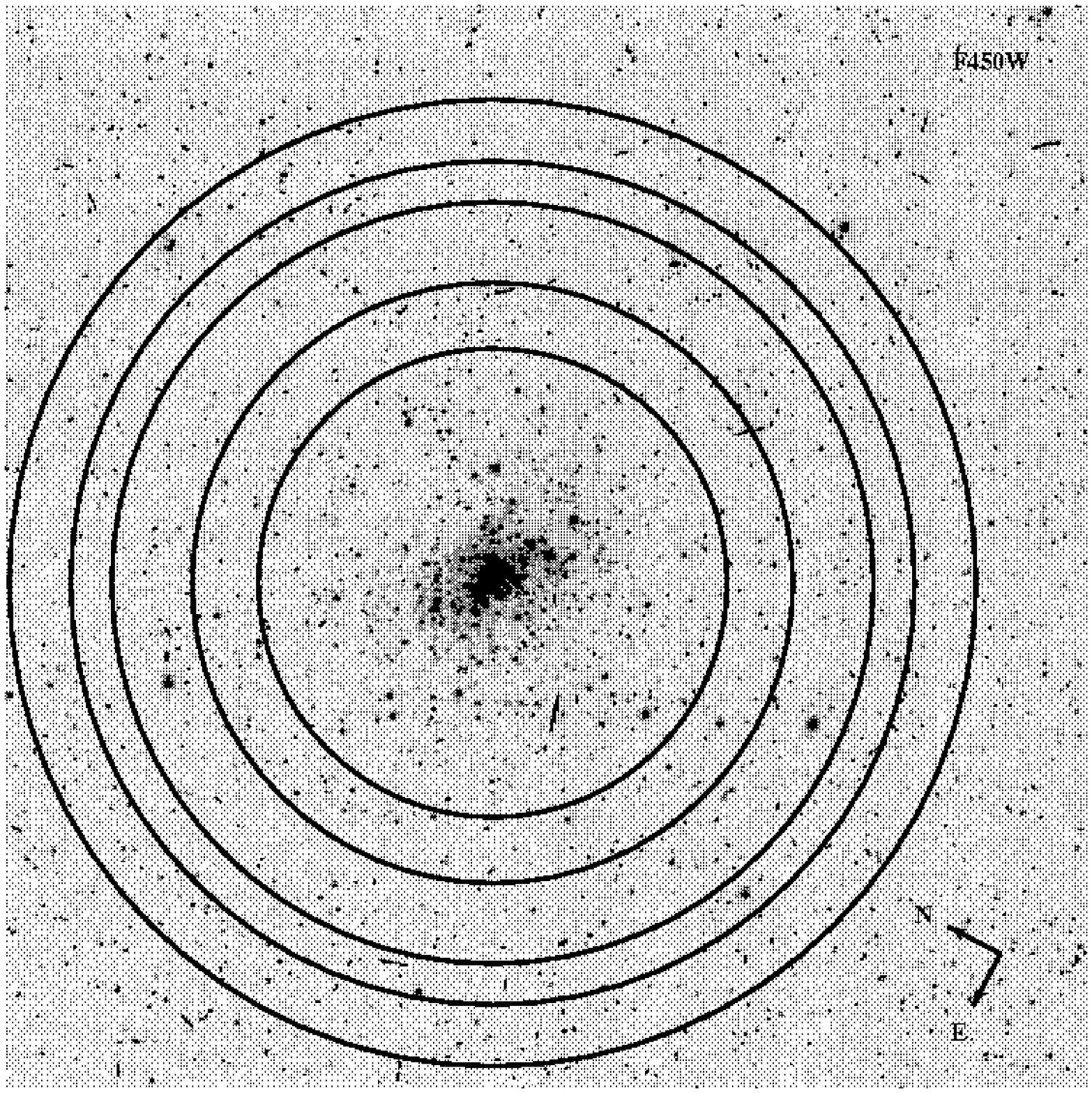}
   \includegraphics[width=9cm]{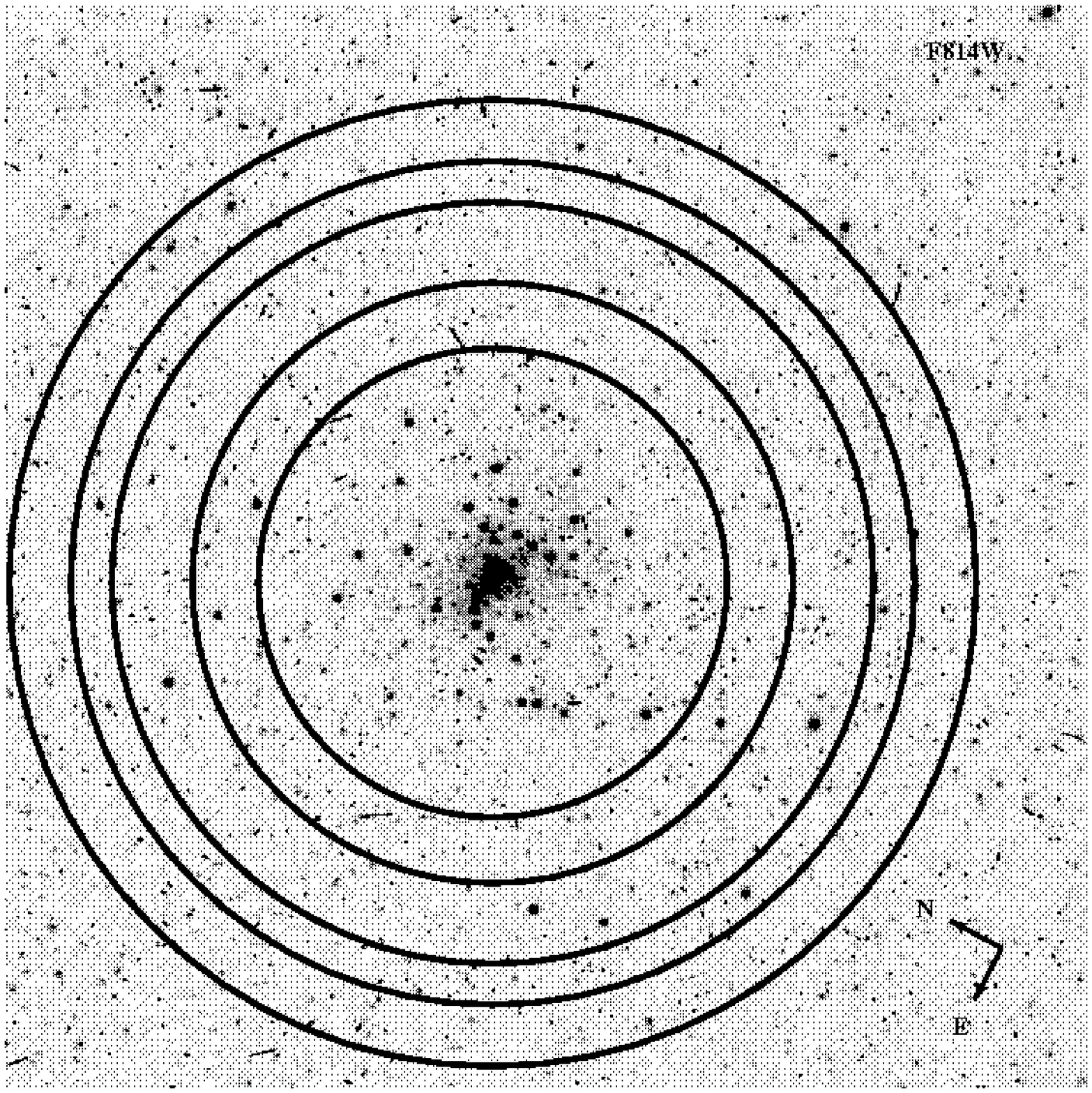}
      \caption{F450W (upper panel) and F814W (lower panel) images of the whole
      PC camera, with VdB0 at the center. 
      The superposed circles have radius 
      r=160, 205, 260, 288 and 330 pixels, from inside out, 
      and mark the edges of the
      annuli whose CMDs are shown in Fig.~\ref{clus}, below.
      The light stripes associated with stars
      in the F450W image are due to the effect of CTE that is particularly 
      strong in this shallow low-background image.}
         \label{ima}
   \end{figure}

VdB0 is located at a projected distance of $R_p= 10.8$ kpc
from the center of M31 to the South-West, just $\sim 4\arcmin$ 
from the major axis of the galaxy (see Tab. ~\ref{table:1}), near the edge 
of one of the most prominent substructures of the M31 disk, 
the so called {\em 10 kpc ring}
(see Hodge \cite{hodge} and Barmby et al. \cite{spitzer}, and references 
therein) and within a the large OB association OB80 (van 
den Bergh \cite{vdbOB}, A80 in Hodge (\cite{atlas}) atlas). Its radial velocity 
($V_r=-567$ km/s, Perrett et al. \cite{perrett}) is in full agreement with 
the rotation curve of the HI disk of M31 (Carignan et al. \cite{rotcur}), 
thus confirming the physical association with the thin disk of 
the parent galaxy (F05). The strong value of the $H_{\beta}$ 
index supports the idea that the cluster is younger than 1 Gyr 
($H_{\beta}=4.3$ \AA,  Perrett et al.~\cite{perrett}\footnote{Note that
Perret's et al. measures refers to B195D, i.e. the ``alter ego'' of VdB0 whose
available coordinates were the most appropriate for the cluster.
In this context, it is interesting to note that, adopting a calibration based 
on old  GCs, Perrett et al. found [Fe/H]=-1.64 for VdB0, from integrated
spectral indices (see F05).}).
The existing estimates of both $V_r$ and  $H_{\beta}$ 
are nicely confirmed by recent high signal-to-noise spectra 
acquired at the Italian Telescopio Nazionale Galileo (S. Galleti, 
private communication). 

With the assumed reddening and distance, the integrated V 
magnitude reported in the RBC (see Tab.~\ref{table:1}) gives an absolute 
magnitude $M_V=-10.03$,
much brighter than any Galactic open 
cluster older than 10 Myr (see Bellazzini et al. \cite{cefa}, and below); 
it appears quite extended and irregular in shape even in ground 
based images. In these ways VdB0 stands out among the members 
of our candidate BLCC sample that are, in general, fainter 
and more compact than it.

\section{Observations and Data Reduction}

Our survey was originally planned for the Advanced Camera for Surveys (ACS) 
but it was performed with the Wide Field and Planetary Camera 2 
(WFPC2)
during cycle 16 because of the  failure of ACS.
For each target of our survey 
we acquired two F450W and two F814W images, all with 400 s 
exposure time and gain $=7 e^-/DN$. 
The pointings were chosen 
to place the main target at the center of the PC ($800\times 800 ~px^2$, 
with pixel scale $0.045$ arcsec/px), while the three WF cameras 
($800\times 800 ~px^2$, with $0.099$ arcsec/px) are supposed to sample 
the surrounding fields. The images of VdB0 discussed here were 
acquired on July 2, 2007. 
The image of the whole WFPC2 mosaic image is shown in Fig.~1. It is clear that
there are substructures and density gradients on the scale of the whole mosaic
image, mainly due to the inclusion of the edges  of the large stellar
association embedding the cluster (A80, Hodge \cite{atlas}). As the overall
stellar density on the  WF2 field is larger than in WF3 and WF4, we make the
conservative choice to adopt the WF2 as our preferred sample of the background
population  that is expected to contaminate the Color Magnitude Diagram of the
cluster, while we will consider the average density over all the WF fields when
we will compute stellar density profiles based on star counts (Sect.~3). 
In the present context, when we speak of ``background population'' we
refer to all the stars belonging to the field of M31 but unrelated to the cluster
we are studying.  
Zoomed views of the PC field in both F450W and F814W passbands are shown 
in Fig.~\ref{ima}.

As the observational material and the degree of crowding are 
essentially the same for all the surveyed fields, we tuned our 
data-reduction strategy to be exactly the same in all cases, to 
maintain the highest degree of homogeneity in the final products 
of the survey. Data reduction has been performed on the pre-reduced 
images provided by STScI, using HSTPHOT\footnote{See
{\tt http://purcell.as.arizona.edu/hstphot/}} (Dolphin \cite{hstphot}), 
a Point Spread Function -fitting package specifically devoted 
to the photometry of WFPC2 data. The package identifies 
the sources above a fixed flux threshold on a stacked image 
and performs photometry on individual frames, and automatically 
applies the correction for the Charge Transfer Efficiency
(CTE, Dolphin \cite{dolcal}).  It then transforms instrumental magnitude to 
the VEGAMAG system (see Holtzman et al. (\cite{holtz}) and Dolphin 
(\cite{dolcal})),  
deals with cosmic-ray hits, and takes also into account 
all the information about image defects that is attached to the 
observational material. We fixed the threshold for the search of 
sources on the images at 3~$\sigma$ above the background. 
HSTPHOT provides as output the magnitudes and positions of the detected 
sources, as well as a number of quality parameters for a suitable 
sample selection, in view of the actual scientific objective one 
has in mind. Here we selected all the sources having valid magnitude 
measurements in both passbands, global quality flag = 1 
(i.e., best measured stars),
{\em crowding} parameter $<0.3$, $\chi^2<2.0$ and $-0.5< sharp<0.5$, 
in both passbands, (see Dolphin \cite{hstphot} for details on the 
parameters). This selection cleans the sample 
from the vast majority of spurious and/or badly measured sources 
without significant loss of information, and it has been found 
to be appropriate for the whole survey.

   \begin{figure*}
   \centering
   \includegraphics[width=16cm]{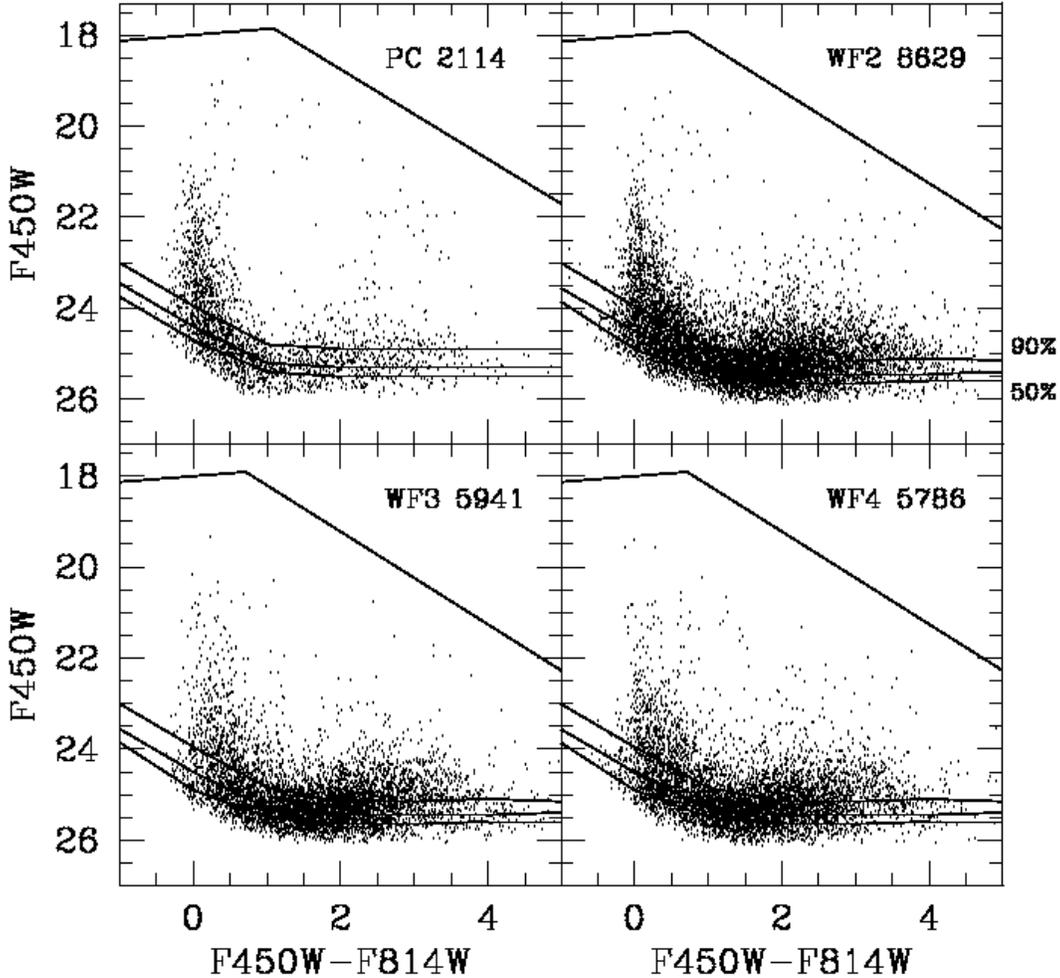}
      \caption{CMD of the fields sampled by the four chips of the WFPC2.
      The number of  stars plotted is reported in the upper left corner of each
      panel. The upper line marks the threshold above which stars saturate
      the intensity scale of the images. The lower lines are CMD loci at the
      same level of completeness, 90\%, 70\% and 50\% from top to bottom,
      respectively (see labels in the WF2 panel).}
         \label{cmd}
   \end{figure*}
%

In Fig.~\ref{cmd} the Color Magnitude Diagrams (CMD) of the fields 
imaged by the four chips of WFPC2 are shown. The threshold 
for the saturation of bright stars and the boundaries at which the 
completeness of the sample reaches 90\%, 70\% and 50\% are also 
shown, as derived from the artificial stars experiments described 
below. As the CMD is quite typical of our survey, it is worthy 
of some general comments while a detailed analysis is deferred 
to Sect. ~3 below. First, our photometry is relatively shallow, 
due the short exposure times of our images; the 50\% completeness 
level is reached at $F450W\simeq 25.5$\footnote{Except for the very crowded region
at the center of the cluster. For 10 px $<r\le$ 50 px, the 50\% completeness level 
is reached at $F450W \ga 23.5$.}.
For the same reason 
our images, and particularly the F450W ones in which the background 
light is very low, are badly affected by CTE (see Fig.~\ref{ima}).
Therefore the accuracy of the absolute and relative photometry 
is not particularly good 
(see, for example, Fig.~\ref{err} and Tab.~\ref{taberr}, below).
In spite of that, the very wide wavelength baseline provided 
by the F450W and F814W filters produces relatively well 
defined sequences in the CMD (compare, for example, with the 
CMD of similar fields obtained by WH01 with the same camera 
and longer exposure times but using F439W and F555W filters). 

All the fields targeted by our survey cross the outer regions of 
the star-forming thin disk of M31 (see F05), and as a consequence, 
in most cases, the most prominent feature of the CMD is the 
nearly vertical plume of young Main Sequence stars that is seen
in Fig.~\ref{cmd} around $F450W-F814W\simeq 0.2$. 
The wide blob of stars at $F450W>24.0$ and  $F450W-F814W\ge 1.5$ is consistent 
with being due to the brightest Red Giants near the tip of 
the Red Giant Branch (RGB) of the old-intermediate population 
that seems to be pervasive in the M31 disk (see Bellazzini et al. 
\cite{mbm31}, and references therein). Red and blue supergiants as well 
as other less-massive evolved stars are likely present at bright 
magnitudes over the whole color range covered by our CMD 
(see Massey \cite{massey}).


\subsection{Artificial stars experiments}

The completeness of the samples and the accuracy in the relative 
photometry are best estimated with extensive sets of artificial 
stars experiments (see Bellazzini et al. \cite{288a,288b} and Tosi et 
al. \cite{tosi} for detailed discussions and references).

   \begin{figure*}
   \centering
   \includegraphics[width=16cm]{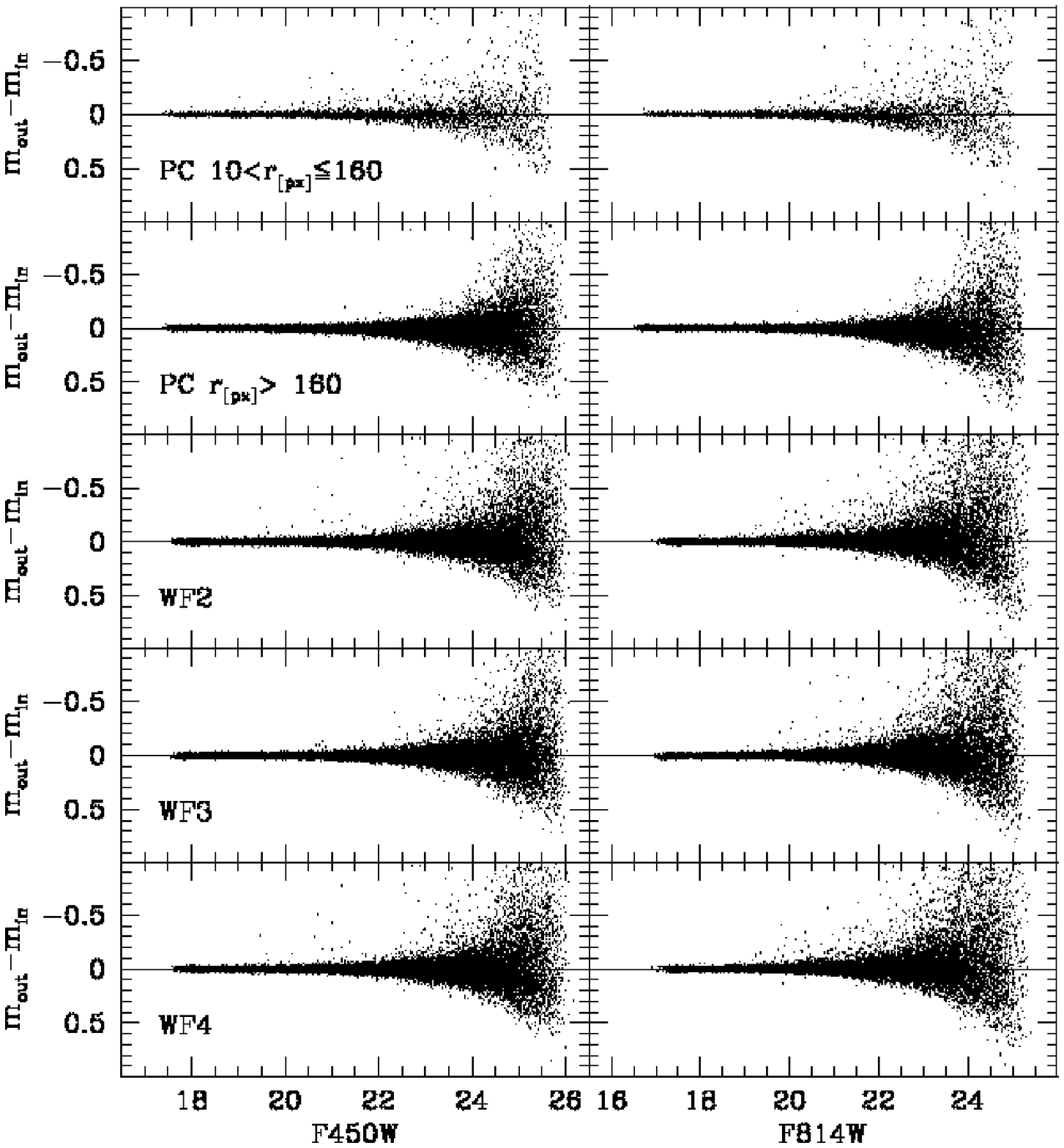}
      \caption{Distributions of the differences between the {\em output} and
      {\em input} magnitudes of artificial stars as a function of F450W (left
      panels) and F814W (right panels) magnitudes, for the PC and WF fields. 
      The top panel displays the
      distributions for the most crowded region of the PC camera, i.e. the one
      containing the cluster. $r_{[px]}$ is the distance from the cluster
      center in PC pixel units, assuming (x,y)=(405,398) as the coordinate of
      the center in the reference frame of the photometric catalogue.
      To make the diagrams more easily readable we plot just a fraction of the
      whole set of artificial stars, i.e. 50000 stars per field, approximately,
      while more than 150000 per field are typically recovered.}
         \label{err}
   \end{figure*}
%

%
\begin{table}
\caption{Uncertainties in the relative photometry from artificial stars
experiments, for 10 px $<r\le$ 160 px, PC field.}             
\label{taberr}      
\centering                          
\begin{tabular}{l c l c}        
\hline\hline                 
F450W & $\sigma^a$ & F814W & $\sigma^a$\\    
\hline 
18.00 &  0.009 & 18.00 &   0.010 \\
18.50 &  0.010 & 18.50 &   0.011 \\
19.00 &  0.010 & 19.00 &   0.012 \\
19.50 &  0.011 & 19.50 &   0.013 \\
20.00 &  0.013 & 20.00 &   0.016 \\
20.50 &  0.016 & 20.50 &   0.020 \\
21.00 &  0.018 & 21.00 &   0.026 \\
21.50 &  0.023 & 21.50 &   0.036 \\
22.00 &  0.029 & 22.00 &   0.050 \\
22.50 &  0.039 & 22.50 &   0.068 \\
23.00 &  0.054 & 23.00 &   0.087 \\
23.50 &  0.076 & 23.50 &   0.138 \\
24.00 &  0.107 & 24.00 &   0.218 \\
24.50 &  0.153 & 24.50 &   0.336 \\
25.00 &  0.241 & 25.00 &   0.377 \\
25.50 &  0.309 & 25.50 &   0.400 \\
\hline                                   
\end{tabular}
\begin{list}{}{}
\item[$^{\mathrm{a}}$] $\sigma$ are $\pm$ 1
      standard deviations after the clipping of outliers at more than $3\sigma$
      from the mean.
\end{list}
\end{table}

HSTPHOT allows easy, fast and fully automated runs of artificial 
stars experiments. Fake stars in a user-selected color range, 
extracted at random from a Luminosity Function (LF) similar to 
the observed one, are added to the original frames one at a time 
to avoid self-crowding (Dolphin, private communication) and 
the photometric reduction is repeated. With the final catalogue 
of {\em input} and {\em output} magnitudes of artificial stars the distribution 
of photometric errors and the completeness of the samples 
can be studied as a function of color and as a function of the 
distance from the center of the cluster under consideration (i.e. as a function 
of crowding). We simulated a total of 728398 artificial stars, 
roughly equally distributed on the four WFPC2 chips. 

Fig.~\ref{err} shows the distributions of the differences between 
the {\em output} and {\em input} magnitudes of artificial stars as a function 
of F450W (left panels) and F814W (right panels) magnitudes, 
providing a direct estimate of the typical uncertainties of 
our relative photometry. The small excess of stars at negative 
$m_{out}-m_{in}$, increasing in number and amplitude of the difference 
for fainter magnitudes, is due to artificial sources that are erroneously 
recovered with a brighter magnitude because they are blended 
with real sources present on the image (see Tosi et al.~\cite{tosi}). 
Even in the most crowded region of the PC that includes the 
cluster (top panels of Fig.~\ref{err}) the effects of blending are not 
particularly severe, at least for relatively bright stars. 
The probability of a star with $F450W\le 23.5$ to have its magnitude decreased 
by more than 0.1(0.2) mag by the combination of blending 
and photometric error is 2.8\%(1.4\%) if its color lies in the 
range $-0.6\le F450W-F814W\le 1.5$ 
and 3.5\%(1.6\%) for $2.0\le F450W-F814W\le 4.0$.
Typical photometric uncertainties 
as a function of magnitude are reported in Table ~\ref{taberr} for the 
innermost region of the PC field, covering most of the cluster 
that is the main subject of the present study.

   \begin{figure}
   \centering
   \includegraphics[width=9cm]{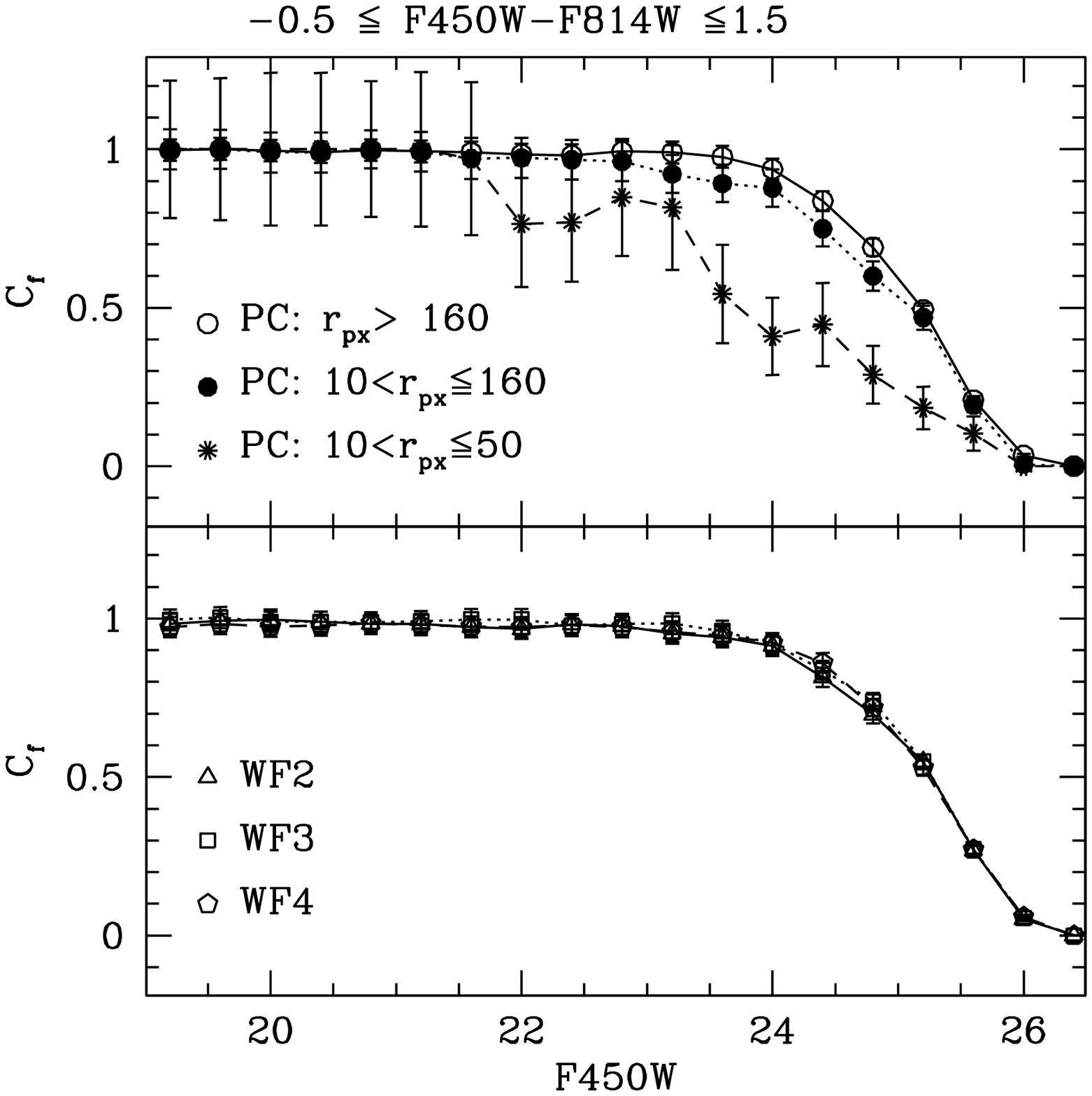}
      \caption{Completeness factor ($C_f$) as a function of F450W magnitude for
      the color range enclosing MS stars. Upper panel: $C_f$ for regions of the
      PC field at different distances from the cluster center. Lower panel:
       $C_f$ for the three WF fields. Note that the three curves are
       indistinguishable within the uncertainties.}
         \label{comple}
   \end{figure}
%

Finally the completeness factors ($C_f$) as a function of magnitude 
for different regions of the PC and for the WF fields are 
shown in Fig.~\ref{comple}, for stars in the wide color range
$-0.6\le F450W-F814W\le 1.5$.
Outside of the innermost region of 
the PC including the cluster, the $C_f$ functions are nearly indistinguishable. 
For $r>$ 50 px the completeness is larger than 80\% 
for $F450W\le 24.0$ and in any case $C_f\simeq 1$ 
(i.e. completeness $\simeq$ 100\%) for $F450W\le 22.0$.

\subsection{Theoretical stellar models}

Most of our inferences about the physical parameters of the stellar 
populations (clusters or field) considered in our survey will 
be obtained from the comparison between the observed CMDs 
and theoretical stellar models, in the form of isochrones or synthetic 
CMDs. The need to have models in the natural photometric 
system in which the observations were obtained (HST/WFPC2 
VEGAMAG) and to have a set of isochrones reaching ages as 
young as 10 Myr led us to chose the set by Girardi et 
al. (\cite{gir02}, hereafter G02), as our reference grid of stellar models. 
In particular we took their HST-color version of the solar-
scaled models by Salasnich et al. (\cite{sala}), with overshooting and 
a simplified TP-AGB evolution, as this set includes 10 Myr old 
isochrones up to super-solar metallicities\footnote{
{\tt http://pleiadi.oapd.inaf.it}}.
In some cases, when 
a particular model is needed, we use the CMD web 
tool\footnote{{\tt http://stev.oapd.inaf.it/~lgirardi/cgi-bin/cmd}}
(Marigo et al. \cite{marigo}),
that allows the on-line computation of models 
from user specified inputs, using the G02 set. 

In some cases, for comparison and/or for special applications, 
we use the BASTI\footnote{http://www.oa-teramo.inaf.it/BASTI/index.php}
database, collecting the theoretical 
models by Pietrinferni et al. (\cite{basti}), and updates. In particular 
BASTI provides a very practical Web Tool to produce 
synthetic CMDs of populations with ages, chemical composition, 
initial mass function, binary fraction ($f_b$) etc. selected by 
the user (Cordier et al. \cite{cordier}), that can be used to compare models 
and observations in term of star counts in different color and 
magnitude ranges (see Fig. ~\ref{ebv}, for an example of application). 
Unfortunately, the models are not provided in the WFPC2 photometric 
system - so theoretical magnitudes have to be transformed 
- and isochrones/synthetic CMDs for ages $< 30$ Myr are not provided; 
for these reasons we didn't adopt the BASTI set as the 
reference for our survey. In the considered range of ages G02 
and BASTI isochrones (with overshooting) provide very similar 
predictions of color and magnitudes, while evolving masses may 
differ by $\sim$ 20\%  
(see also Gallart, Zoccali \& Aparicio \cite{galla}).

\subsection{Reddening and Distance}

\begin{figure}
   \centering
   \includegraphics[width=9cm]{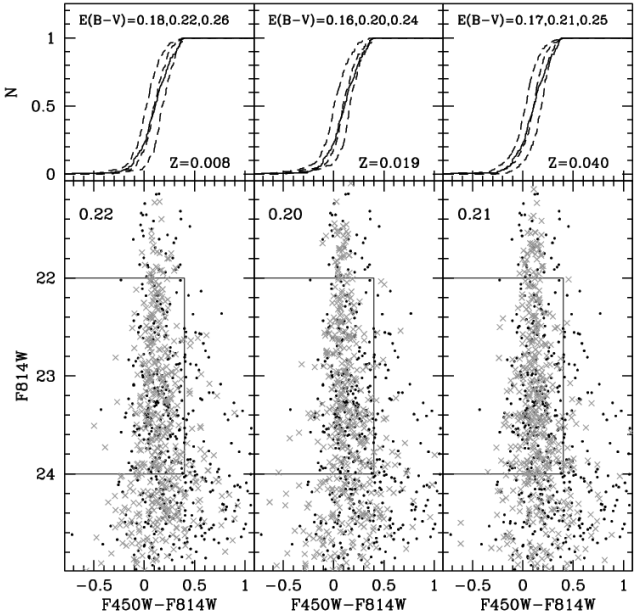}
      \caption{The observed CMD of VdB0 (black dots, only stars with $r\le
      160$ px) is compared with the
      synthetic CMD (grey $\times$ symbols) of 30 Myr old, $f_b=$50\% 
      populations having $Z=0.008$ (left panel), $Z=0.019$ (middle panel),
      and $Z=0.040$, obtained from the BASTI webtool 
      (Cordier et al. \cite{cordier}), transformed 
      to WFPC2-VEGAMAG with Dolphin (\cite{dolcal}) equations, 
      and corrected for photometric errors and completeness 
      according to the results of our artificial stars experiments.
      The thin lines enclose the selection box in which the cumulative color
      distributions shown in the upper panels have been obtained, focusing on
      the blue edge of the Main Sequence. 
      In these panels the observed color distribution (continuous line) is
      compared to the distributions of the synthetic sample of the adopted
      metallicity for
      three different assumptions on the reddening value (dashed lines),
      reported in the upper label. The middle value corresponds to the
      distribution that best fits the observations and is also reported in the
      upper left corner of the CMDs. Note the very weak dependence of the
      reddening estimate on the metallicity of the adopted model.}
         \label{ebv}
   \end{figure}

To correct for the effects of interstellar extinction and reddening 
we will always adopt the relations 
$A_{F450W}=4.015E(B-V)$ and $A_{F814}=1.948E(B-V)$,
as reported by Schlegel, Finkbeiner \& 
Davis (\cite{dirbe}). As our clusters are embedded in the structured 
dusty disk of M31 it does not seem appropriate to assume a 
unique value of reddening for all of them; the typical reddening 
value attributed to Galactic dust toward M31 ranges from $E(B-V)=0.06$
(Schlegel et al. \cite{dirbe}) to $E(B-V)\simeq 0.11$ (see Galleti~\cite{rbc}, 
and references therein), 
but it is likely that our clusters are more reddened than this 
(Barmby et al.~\cite{barm00}; Fan et al.~\cite{fan}). To get an 
estimate of the reddening affecting the clusters in our survey we 
compare theoretical models (isochrones and synthetic CMDs) to 
the observed MS in the range $22.0\la F450W\la 24.0$.
In this range, corresponding to absolute magnitudes $-3.0\la M_{F450W}\la 0.0$,
the color of the MS is only weakly sensitive to metallicity 
and various sets of theoretical models provide very consistent 
predictions. An example of our analysis is presented 
in Fig.~\ref{ebv}, where we compare the color distribution at the blue edge of
the MS of the observed sample and of synthetic samples (from the BASTI webtool)
of different metallicities, adopting different reddening values. 
The comparisons confirm that the sensitivity to metallicity of the
reddening estimate is very weak, as expected.
In the case of VdB0 we obtain $E(B-V)=0.2 \pm 0.03$ with 
this method, and we will always adopt this value below.

In the following and for the whole survey we adopt $(m-M)_0=24.47\pm 0.07$
as the distance modulus of all the considered populations, from McConnachie 
et al. (\cite{dista}), corresponding to an heliocentric distance $D = 783$ kpc.
At this distance $1\arcsec$ corresponds to 3.8~pc, $1\arcmin$ to 228~pc.

\subsection{Accessible age range}
 
\begin{figure}
   \centering
   \includegraphics[width=9cm]{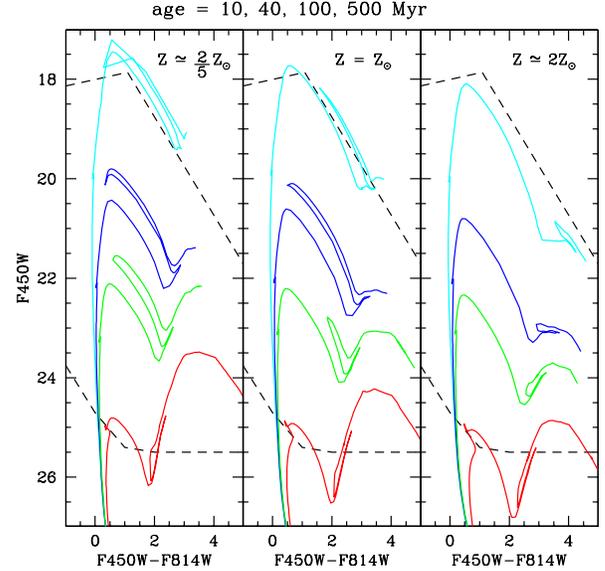}
      \caption{Isochrones of different ages and metal content are plotted on
      the ``visibility window'' of our CMDs, enclosed on the bright side by the
      saturation limits and on the faint side by the $C_f=50$\% line 
      (long-dashed lines). 
      The continuous curves are isochrones from the G02 set; ages and
      metallicities are indicated in the figure.}
         \label{eta}
   \end{figure}

As the degree of crowding of all the surveyed fields is quite similar 
and the observational set-up is identical in all cases, the saturation 
limit and the $C_f=0.50$ 
limit reported in the CMDs of 
Fig.~\ref{cmd} can be considered representative of the typical CMD window 
that is accessible with the survey data. In Fig.~\ref{eta} we compare 
isochrones of different ages and metallicities with this window 
to have an idea of the age range in which we can obtain reasonable 
age estimates for the considered clusters from the luminosity of 
their Turn Off 
(TO) points and/or from the distribution of their 
Super Giant populations.

%

In the metallicity range that is most likely to enclose the disk 
populations (we are considering $\frac{2}{5} Z_{\sun}\la Z\la 2 Z_{\sun}$)
we can detect 
the TO point of clusters roughly ranging from 10 to 500 Myr old. 
As the only BLCCs for which a direct CMD-based age estimate has been obtained
are 60-160 Myr old (WH01), the age sensitivity of the survey 
seems rather appropriate; however clusters in the age range 0.5 - 2 
Gyr may prove very difficult to age date with our data. For the 
oldest populations (age $\ga 2$ Gyr) we can hope to detect just the 
tip of the RGB, as shown by the age=12 Gyr isochrones plotted 
as thick lines in Fig.~\ref{clus}, below.

\section{The CMD and structure of the cluster VdB0}

\subsection{Distribution of resolved stars}

To identify the stellar population of the cluster as securely as possible,  it
is useful to have an idea of the surface density distribution of its resolved
stars.  In the present context we are interested only in defining the
characteristic size of the region dominated by cluster stars, in order to select
samples of likely cluster members by radius (see Sect.~3.4 for a detailed
analysis of the light profiles).

   \begin{figure}
   \centering
   \includegraphics[width=9cm]{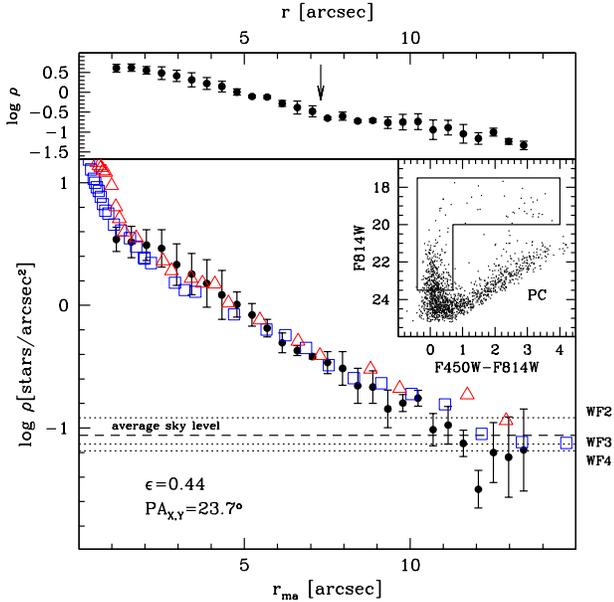}
      \caption{Upper panel: Background-subtracted surface density profile of 
      VdB0 computed by counting stars on circular concentric annuli around the 
      center of the cluster. The arrow marks the radius where a sudden change
      of slope in the profile appears, at $r\simeq 160$ px $=7.3\arcsec$.
      Lower panel: Background-subtracted profile from star-counts (filled
      circles with errorbars) converted to a major-axis profile, adopting the
      reported values of PA and $\epsilon$. Open symbols are the corresponding
      light profiles described in Sect.~3.4, squares for the F450W image 
      and triangles for F814W, vertically shifted by an arbitrary 
      normalization to match the star counts at $r_{ma}>3\arcsec$. 
      The dotted lines mark the average surface density in each of the WF
      cameras, the dashed line is the average of the three, which was in the end
      adopted as the background value to subtract to star-count profiles.   
      Only stars within the L-shaped box plotted in the CMD in the upper right
      corner of the lower panel are selected for star counts, 
      as probable cluster members.}
         \label{prof}
   \end{figure}
%

Stars were selected on the CMD from the box shown in the diagram enclosed in the
lower panel of Fig.~\ref{prof}. The box is
expected to pick up the best-measured MS and SG stars typical of the
cluster population, while excluding populations that are clearly not associated
with the cluster, such as the much older stars around the tip of the RGB.
For $r\lesssim 3\arcsec$ star counts are significantly affected by 
radially varying incompleteness in the  range of magnitudes considered.
Beyond this limit the
degree of completeness is fairly high and essentially constant with radius  
(see Fig.~\ref{comple}, above), hence the derived profile should be reliable.

In the upper panel of Fig.~\ref{prof} we show the surface density profile 
obtained by counting stars on circular annuli centered on the cluster center.
The observed profile displays an obvious break at $r\simeq
7.3\arcsec$, where it begins to decline with a gentler slope 
out to $r\sim 14\arcsec$. 
The break in the profile may reflect an 
{\em inner core} + {\em outer corona} structure of VdB0, which 
is typical of Galactic Open Clusters (see Kubiak et al. \cite{tom2}, 
Kharchenko et al. \cite{kar}, Mackey \& Gilmore \cite{mack}, 
Elson et al. \cite{els}, and references
therein), or it may be ---at least partially-- due to the elongated distribution
of the cluster stars unaccounted for by our adoption of circular annuli.
To investigate this possibility we transformed the radial coordinate of each
star ($r$) into a major-axis radius ($r_{ma}$) defined as

\begin{equation}
r_{ma}=\sqrt{X_r^2+\left({1\over{\left(1-\epsilon\right)}}Y_r\right)^2}
\label{rma}
\end{equation}

where

\begin{equation} 
X_r=(X-X_0)cos(PA_{X,Y})+(Y-Y_0)sin(PA_{X,Y})
\end{equation}
\begin{equation} 
Y_r=-(X-X_0)sin(PA_{X,Y})+(Y-Y_0)cos(PA_{X,Y})
\end{equation}

\noindent
and ($X_0,Y_0$) are the coordinate of the center of the cluster, 
$\epsilon=1-b/a$, is the ellipticity, where $a$ and $b$ are the semi-major
and semi-minor axis, respectively, and $PA_{X,Y}$ is the position angle measured
from the X axis toward the Y axis. Both $\epsilon$ and $PA_{X,Y}$ are taken 
(or easily derived, in the case of $PA_{X,Y}$) from the results of 
the analysis of the light distribution presented in Sect.~3.4, below. 
Eq.~\ref{rma} has been adapted to our case from Eq.~4 by Martin et al.
(\cite{mart}).

The ellipticity-corrected major axis profile is plotted in the lower panel of
Fig.~\ref{prof}, and it clearly shows that the change of slope in the original 
profile was an artifact due to the inadequacy of the assumption of 
circular symmetry. The result is supported by the good match between the
star-counts profile and the light profiles (from Sect.~3.4) 
over the large radial range where they can be compared ($r> 3\arcsec$). 

It is interesting to note that the cluster
profile appears to extend to remarkably large distances from the center, out to
$\simeq 15\arcsec \simeq 57$ pc.
As the process of profile analysis described in Sect.~3.4 includes also 
the fitting of King (\cite{k66}, hereafter K66) models, 
it is interesting to note that the limiting radius of the K66 models that 
best fits the surface brightness profiles is also $r_t\simeq 15\arcsec$, 
thus supporting the conclusion that the cluster is very extended.

The elongated shape of the cluster will be taken into account in the
detailed analysis of the profiles of Sect.~3.4. 
For present purposes it is sufficient to  conclude that most of the cluster 
stars are  enclosed within a (circular) 
radius of $7.3\arcsec$ (160~px) from the center. 
We take this as a  reference radius for the following analysis of the CMD,
as it allows a very simple radial selection, 
remembering that some cluster members are also present at larger radii.

   \begin{figure*}
   \centering
   \includegraphics[width=\textwidth]{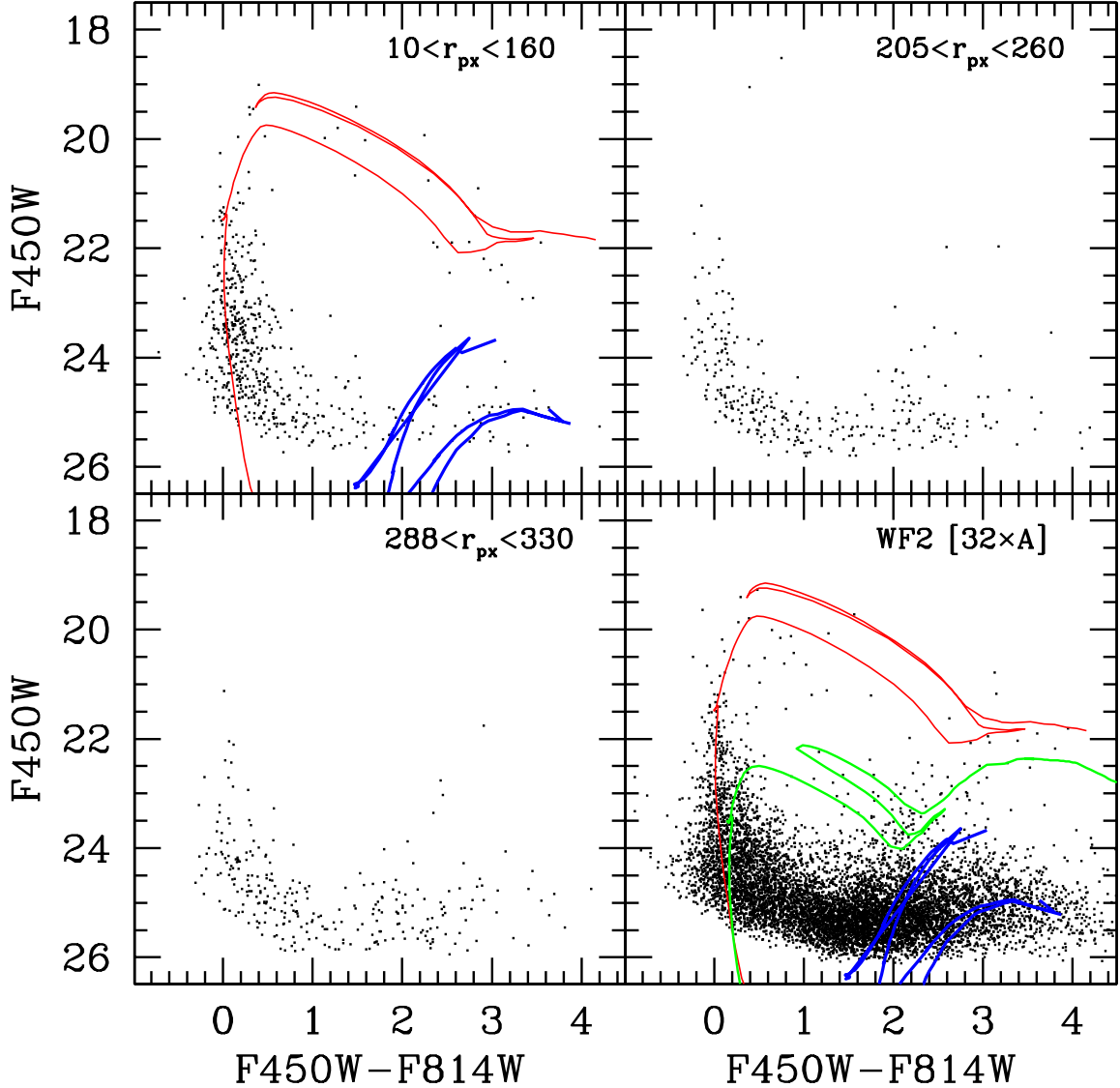}
      \caption{CMDs of different circular annuli around the center of VdB0 in the
      PC field (see Fig.~\ref{ima}, above), 
      all having the same area, (upper panels and lower left panel) 
      and of the whole WF2 field, whose area is 32 times that of the PC annuli
      (lower right panel). 
      The thin line is a $Z=Z_{\sun}$ isochrone of 
      age 25 Myr; the heavy lines at F450W $\le 24.0$ 
      are 12 Gyr old 
      isochrones of metallicity $Z=6\times10^{-4}$ and
      $Z=6\times10^{-3}$, from blue to red, respectively. 
      The additional isochrone plotted in the lower right panel has
      $Z=0.008$ and age 125 Myr. All the isochrones are from G02.}
         \label{clus}
   \end{figure*}
%

The upper left panel of Fig.~\ref{clus} shows the CMD of stars within 
$10 \le r< 160$~px,  an annulus that, as stated earlier, 
should be dominated by cluster stars.
The innermost $r\le 10$ px region has been excluded because of severe
incompleteness.
A main sequence with a TO around $F450W\sim 21.5$ is the most populated
branch of the diagram, with a blue edge at $F450W-F814W\simeq 0.0$. Blue and
red supergiants (BSGs, RSGs) are clearly identified, spanning a large color 
range ($0.0 \lesssim F450W-F814W \lesssim 3.6$~mag). A 25~Myr isochrone of solar
metallicity (from the G02 set) seems to provide a satisfactory fit to the MS and
to the sizable luminosity range spanned by supergiants, suggesting an extended
Blue Loop phase (see Williams \& Hodge \cite{will}). The color of the
reddest supergiants is not fully reproduced (a long standing and
not-so-critical problem of theoretical models, see Massey \cite{masrev}).
An handful of field RGB stars (at F450W$\ge 24.0$ and F450W-F814W$\ga 2.0$) 
is the only population identified in this inner annulus which is clearly not 
associated with the cluster.

The upper right and lower left panels of Fig.~\ref{clus} shows the CMD of outer
annuli of the PC field with the same area as the 10 px $<r<$ 160 px annulus.
Even if these fields still contain some cluster members, their stellar mix should be
fairly representative of the surrounding field population (compare with the WF2
CMD shown in the lower right panel). The comparison of the innermost annulus
with the outer two of the same area shows that the supergiant population is
characteristic of the cluster and is much less frequent in the field, suggesting
an older average age of the field population with respect to the cluster. The
comparison between the morphologies of the MS is consistent this view.
The lower right panel of the figure shows the CMD of a WF2 field whose area is
32 times that of the annuli described above. The larger sampled area provides
a clearer picture of the population mix of the M31 disk in the surroundings
of VdB0. While MS and evolved stars of age (mass) similar to that encountered 
in the cluster are present, the majority of the stars seem to have ages greater
than 100~Myr. In particular the evolved stars at $F450W-F814W \gtrsim 2.0$ and
$F450W \lesssim 24.0$ that are well fitted by the over-plotted 125~Myr, $Z=0.008$
isochrone are not seen in the 10 px $<r<$ 160 px annulus.

The CMD of the cluster (innermost annulus) is very similar to that of rich Large
Magellanic Cloud clusters of age $\sim$ 30-50 Myr, such as NGC~1711 (Sagar et al.
\cite{sagar}) and, in particular, NGC~1850 (Vallenari et al. \cite{valle},
Gilmozzi et al. \cite{gil}).

\subsection{Supergiant Stars}

   \begin{figure*}
   \centering
   \includegraphics[width=16cm]{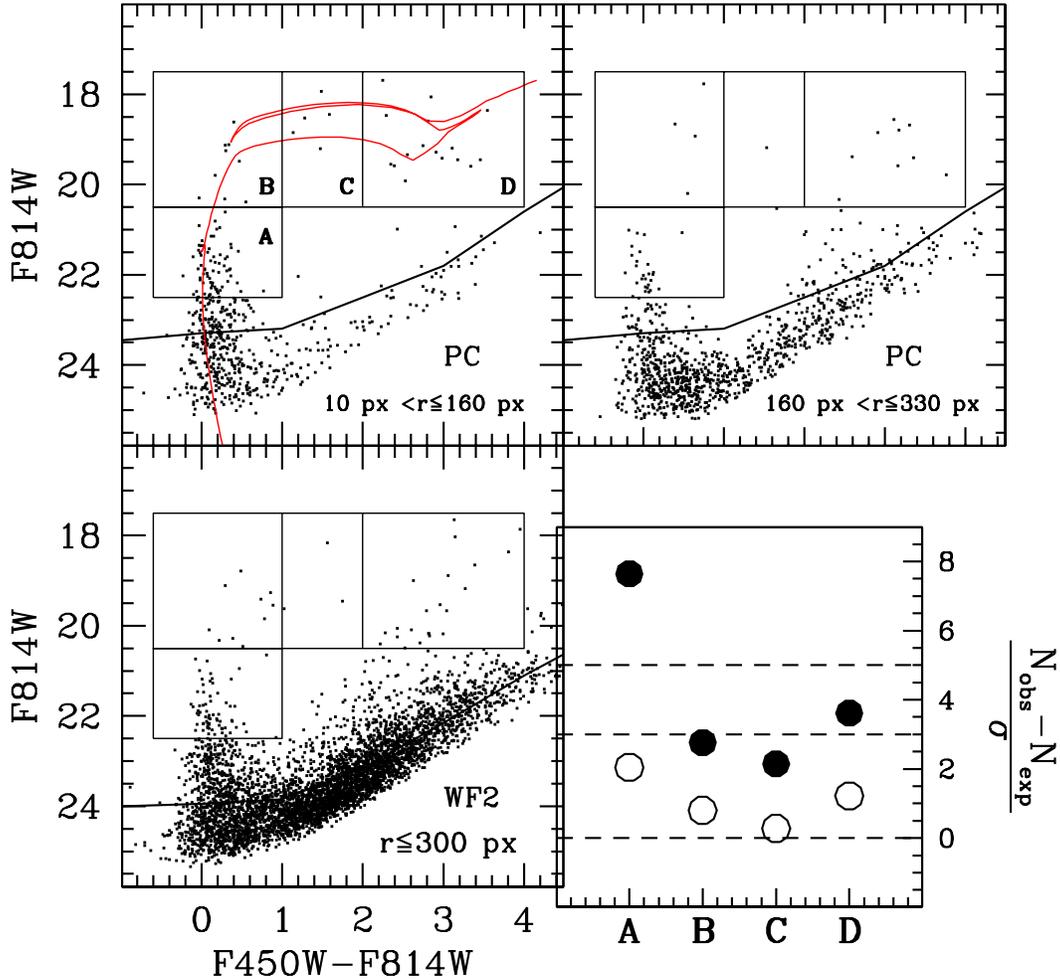}
      \caption{CMDs of different annuli around the center of VdB0 in the
      PC field (upper panels) and of a large area in the WF2 field (lower
      left panel), expected to sample the surrounding ``field'' 
      population. An isochrone of $Z=Z_{\sun}$ and age 25 Myr is superposed
      on the upper left CMDs, as a reference. 
      The $C_f=0.90$ line is reported and
      a raster of labeled boxes is also over-plotted.
      The lower right panel reports the background-subtracted star counts 
      (see Tab.~\ref{table:2}) in the
      various boxes, in units of $\sigma$, for the inner ($r\le 160$ px,
      filled circles) and outer (160 px $< r\le$ 330 px, open circles) annuli. 
      Zero, three and five $\sigma$ levels are marked by dashed horizontal 
      lines.
      }
         \label{rsg}
   \end{figure*}
%

The analysis illustrated in Fig.~\ref{rsg} and reported in Table~\ref{table:2}
quantitatively demonstrates the presence of a significant overabundance of
supergiants in the cluster with respect to the surrounding field. We counted
stars in the different boxes on the CMDs shown in Fig.~\ref{rsg}, sampling the
upper MS (box A) and supergiants of blue (B), intermediate (C) and red (D)
colors. The counts obtained in the $r \le 160$ px and 160 px$< r <$ 330~px annuli
are compared with those expected from the field population, computed by
rescaling the observed counts in the WF2 field by the ratio of the sampled areas.
The lower right panel shows that in the $r \le 160$~px
annulus a clear excess of stars is present in all of the boxes considered. 
The excess of bright MS stars is very
significant and the excess of RSGs is above the 3$\sigma$ level. 
Even if the low number of stars prevents the detection of significant
excesses, the 160 px$< r<$ 330 px annulus shows some excess with respect to 
the field in all of the considered boxes, in agreement with the results of
Fig.~\ref{prof}.

%
\begin{table*}
\caption{Star counts in the CMD boxes defined in Fig.~\ref{rsg}.
Box A samples the upper MS, boxes B, C, and D samples SG stars of blue,
intermediate and red colors, respectively. $N_{exp}$ is the number of stars
expected in a given box from the field population, computed by
rescaling the observed counts in the WF2 field by the ratio of the sampled 
areas. The ratio between the area of the considered field (annulus) and the area
of the WF2 field (used as representative of the field population) is reported in
the last column.}             
\label{table:2}      
\centering          
\begin{tabular}{c c c c c c c c c c}     
\hline\hline       
Field & \multicolumn{2}{c}{Box {\bf A}} &\multicolumn{2}{c}{Box {\bf B}} & 
\multicolumn{2}{c}{Box {\bf C}} & \multicolumn{2}{c}{Box {\bf D}} & Area$_{field}$/Area$_{WF2}$\\ 
      & $N_{obs}$ & $N_{exp}$ & $N_{obs}$ & $N_{exp}$ & $N_{obs}$ & $N_{exp}$& $N_{obs}$ & $N_{exp}$ &\\
\hline                    
PC: 10 px $<r\le$ 160 px    & 68  & $4.8\pm0.6$ & 9 & $0.7\pm 0.2$ & 5 & $0.2\pm 0.1$ & 16 & $1.5\pm 0.3$ & 0.0708\\  
PC: 160 px $<r\le$ 330 px   & 27  &$15.7\pm1.9$ & 4 & $2.3\pm 0.7$ & 1 & $0.7\pm 0.4$ &  9 & $5.1\pm 1.1$ & 0.2314\\  
WF2: $r\le$ 300 px                & 68  &  ---        &10 & ---          & 3 & ---          & 22 & ---          & 1.0000\\  
\hline                  
\end{tabular}
\end{table*}
%

The total background-subtracted number of RSGs attributable to VdB0 is
$\simeq 18$. The true number is likely larger than this, as some RSGs are likely to 
reside in the innermost $r \le 10$~px, which are not included in the present 
analysis as they are not well resolved in our images. 
According to Figer (\cite{figer}) a richer harvest of RSGs
is observed in only one known YMC of the Milky Way, 
RSGC2, with twenty-six RSG stars. 
RSGC1 has fourteen, while other young clusters listed by Figer have less than 
five. RSGC2 is reported to have an age $\le 21$~Myr, 
RSGC1 has age $\le 14$~Myr, and all the other clusters listed by Figer have
ages $\le 7$~Myr, i.e. younger than VdB0 (see below).
As noted above, some rich clusters of similar age are known in the LMC (Vallenari
et al. \cite{valle}, Brocato et al. \cite{broc}), but even there RSGs are not 
present in large numbers.

\subsection{Age and metallicity}

\begin{figure*}
   \centering
   \includegraphics[width=16cm]{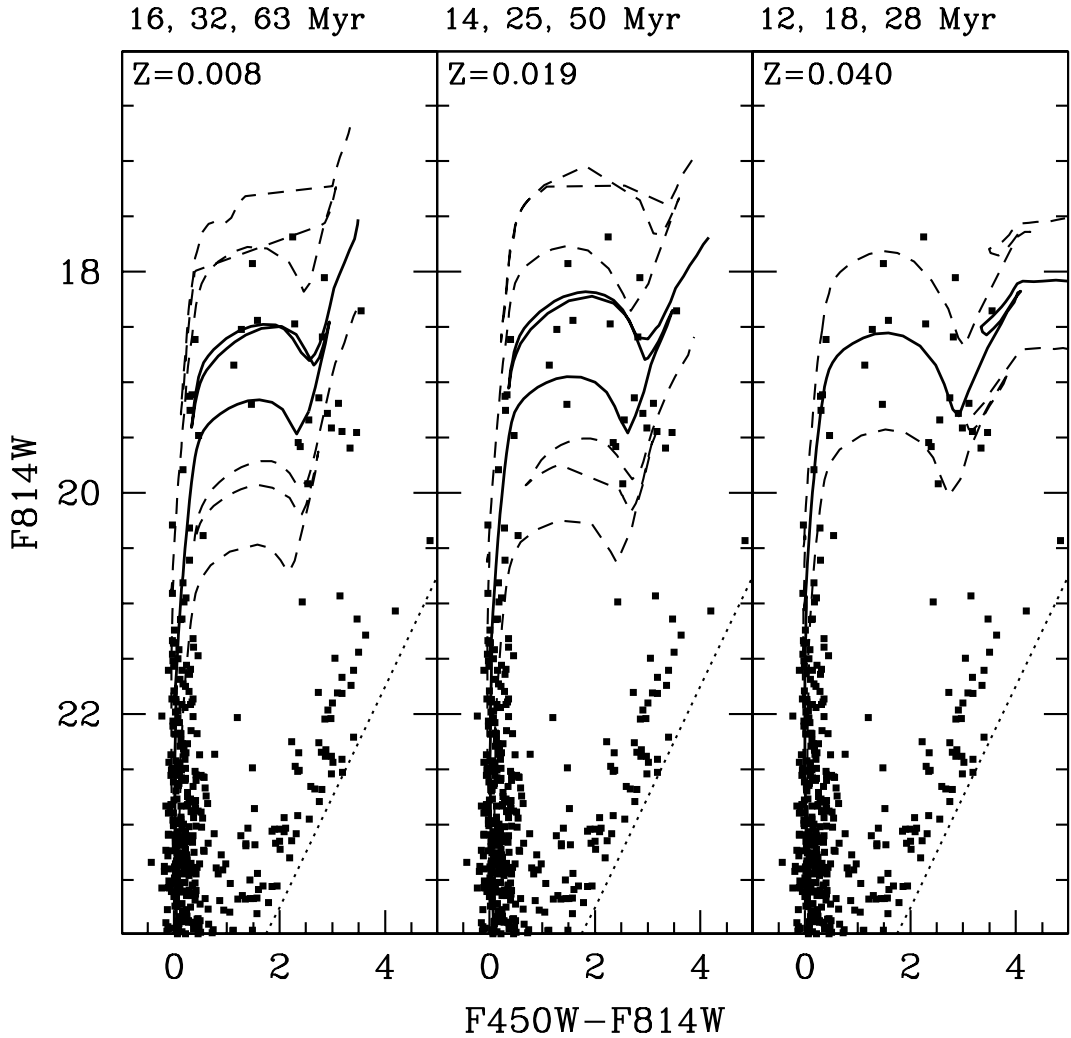}
      \caption{Age estimates for VdB0 for different assumptions about the total
      metallicity (Z). Isochrones from the G02 set are compared to the CMD of
      the cluster (10~px $ < r <$ 160~px).
      The best-fit isochrone is plotted as a thick continuous
      line while the dashed isochrones bracket the upper and lower limits on age.
      The ages and metallicities of the adopted isochrones are reported in each
      panel. The dotted lines mark the limiting magnitude as a function of
      color: the diagonal plume of stars just above the lines (with
      F450W-F814W$>$ 1.5) is populated by likely RGB and AGB field stars, 
      not associated with the cluster.}
         \label{agei}
   \end{figure*}

Having fixed the amount of reddening and the distance modulus to the cluster, 
we obtain an age estimate and an indication of the metallicity 
by comparison with isochrones from the G02 set, following the approach used
by WH01. In
Fig.~\ref{agei} we present a comparison with isochrones of various metallicities
in the range $\frac{2}{5} Z_{\sun}\la Z\la 2 Z_{\sun}$. In all the panels, the
isochrone that is judged (by eye) to provide the best-fit to the observed 
CMD is plotted
as a continuous line. Dashed lines correspond to isochrones providing
strong upper and lower limits to the age estimates, which serve as conservative
estimates of the associated uncertainties.

%

The first very basic conclusion to be drawn from the reported 
upper/lower limits, is that, independent of
the adopted metallicity, the age of VdB0 must be 
within the relatively narrow range from 12 to 63 Myr.

\begin{figure}
   \centering
   \includegraphics[width=8cm]{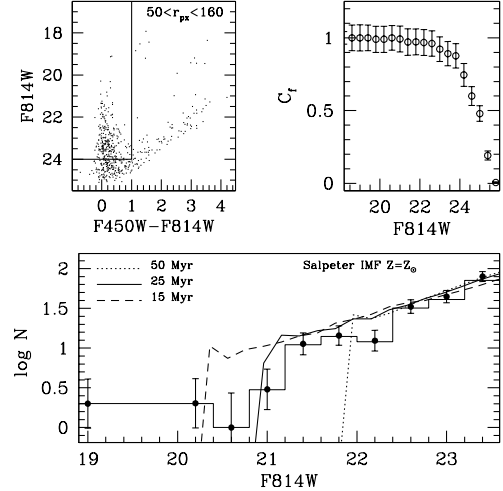}
      \caption{Comparison of the observed LF with theoretical models from the
      G02 suite.
      Upper left panel: CMD of VDB0 with overplotted the box adopted
      to select the sample of stars to be included in the LF. 
      The considered radial range avoids
      the innermost region where the completeness displays significant radial
      variations in the range of magnitudes considered. Upper right panel:
      completeness as a function of magnitude for the  color
      and radial range considered.      
      Lower panel: the observed LF (before completeness correction
      = histogram; corrected for completeness = filled circles with error bars )
      is compared with models of different ages. Note the good fit of the drop
      at $F814W\simeq 21.0$ achieved by the age=25~Myr model. The theoretical LF
      have been arbitrarily normalized to best match the
      three faintest observed points .}
         \label{lf2}
   \end{figure}

The wide range in magnitude covered by supergiant stars strongly indicates
the presence of a wide {\em blue loop} (Massey \cite{masrev}). The
super-solar isochrones clearly lack this feature, hence can likely be excluded
as a possible solution. The larger range of color and magnitude covered by the
$Z=Z_{\sun}$ isochrone in the blue loop phase seems to provide a slightly 
better description of the CMD, compared to the $Z=0.008$ case. 
We produced a set of synthetic CMDs
for populations having $Z=0.008,0.019,0.04$, age~30 Myr and 50 Myr, 
Kroupa (\cite{kroupa}),
Salpeter (\cite{salp}) and $N(m)\propto m^{-1.35}$
Initial Mass Functions\footnote{Salpeter's IMF has $N(m)\propto m^{-2.35}$;
Kroupa's IMF has $N(m)\propto m^{-2.3}$ for $M\ge 0.5 M_{\sun}$, and
$N(m)\propto m^{-1.3}$ for $M<0.5 M_{\sun}$} (IMF), 
using the dedicated Web Tool provided by
the BASTI team. After applying the appropriate distance modulus and reddening
correction and transforming to the HST VEGAMAG system using the
transformations by Dolphin (\cite{dolcal}), we computed a Blue to Red Supergiant 
ratio defined as the ratio of stars having $F814W<20.0$ and $F450W-F814W<2.0$
(B) or $F450W-F814W>2.0$ (R). Independent of age and IMF,  all the $Z=0.008$ models have
$B/R\le 0.26$ ($B/R\le 0.02$~mag for age = 30~Myr), while the observed number
is $B/R=0.60\pm 0.27$. The $Z=0.04$ models have $0.15\le B/R\le 0.52$, while
the solar models have $0.61\le B/R\le 1.17$. Therefore, the color distribution
of SGs provides further quantitative support to the conclusion that the
metallicity of VdB0 is nearly solar. Adopting $Z=Z_{\sun}$ as our best
estimate for the cluster metallicity, the age may be more quantitatively
constrained by the comparison of the observed MS Luminosity Function with 
those predicted by models of various ages. Fig.~\ref{lf2} 
clearly shows that an age=25~ Myr model provides the best-fit to the observed
drop in the star counts at $F814W\simeq 21.0$. The result is well reproduced
also if a Kroupa IMF is adopted.

%
%

Our age estimate is not expected to depend critically on the set of
theoretical models adopted. In their thorough comparison, Gallart, Zoccali \&
Aparicio (\cite{galla}) showed that there is reasonably good agreement between
all the theoretical isochrones they considered in this range of ages (i.e.
 $\le 100$~Myr), if stellar models with core overshooting are assumed.
Our own (limited) set of experiments with Pietrinferni et al. (\cite{basti})
models also supports this conclusion.
A few tests with a set of isochrones adopting the canonical treatment of
convection  (from Pietrinferni et al. \cite{basti}) has shown that the adoption
of such models would lead to younger age estimates, by a factor of
$\sim \frac{3}{5}$, compared to models including overshooting.

Given all the above,
{\em we adopt $Z=Z_{\sun}$ as our best guess for the cluster metallicity, and
25~Myr 
as our best estimate of its age} (see Table~\ref{table:3}). 
The mass of the stars at the TO of the best-fit isochrone 
is $M_{TO}=9.7~M_{\sun}$. 

This relatively rough age estimate is sufficient for our purposes. 
Our final aim is to place the cluster into a log(Age)
versus absolute integrated magnitude diagram such as that shown in Fig.~\ref{vdb0},
below (see also Bellazzini et al.~\cite{cefa}, hereafter B08, 
and references therein), to
compare its stellar mass with that of Galactic open clusters of similar ages.
The uncertainties reported here as the adopted upper and lower limits to
the age estimates correspond to $\la \pm 0.3$ dex in log(Age). These imply 
relatively small changes in the final estimate of the total stellar mass 
(a factor of $\la 2$); the mass estimate also depends relatively weakly 
on the assumed
IMF - see below - and very weakly on the metallicity, at least in the range
considered here, see B08). 


\subsection{Integrated photometry, surface brightness profile and structural parameters}
\label{sb}

Surface-brightness profile-fitting was carried out using methods
similar to those of Barmby et al. (\cite{bar07}). A more detailed description
and the results of profile-fitting for the full cluster sample
will be presented in Barmby et al. (2009, in prep.). Briefly,
the two PC images in each filter were combined with the
STScI Multidrizzle software. 
Intensity profiles were measured using the ellipse fitting routine in IRAF, 
on logarithmically-spaced isophotes centered on the intensity peaks
of the clusters.
The isophotal profiles were `circularized' by converting the semi-major
axes $a$ of the ellipses to effective radii $R_{\rm eff}=\sqrt{a(1-\epsilon)}$,
converted to electrons~s$^{-1}$~arcsec$^{-2}$ by multiplying by 
$({\rm 1 pixel}/0.0455^{\prime\prime})^2=483.033$ and then to intensity in 
$L_{\sun}$~pc$^{-2}$ by multiplying by 14.276 and 6.746
for F450W and F814W, respectively\footnote{This conversion
assumes DN zeropoints of $Z_{450}=21.884, Z_{814}=21.528$,
a gain of 7 electrons~DN$^{-1}$, and
$M_{\rm \sun, F450W}=5.31$ and $M_{\rm \sun, F814W}=4.14$.}.
The mean ellipticity and position angle obtained from the 
analysis of F450W and F814W images are very similar. 
For this reason we take their average as our best values, 
$\epsilon=0.44$ and PA=$45.5\degr$, measured from North toward East.
The available prescription for correcting WFPC2 photometry for CTE 
effects deals only with photometry of point sources, not semi-resolved
objects such as extragalactic star clusters; accordingly, no CTE
corrections were made to the profiles.

\begin{figure}
   \centering
   \includegraphics[width=8cm]{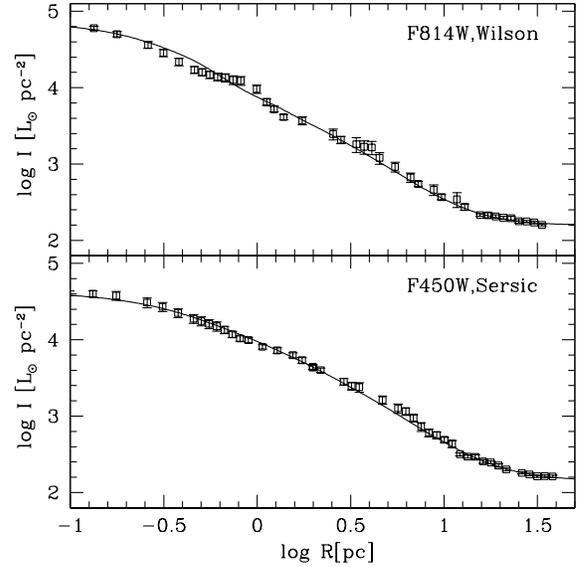}
      \caption{Intensity profiles from surface photometry in circular annuli
      from the F814W image (upper panel) and for the F450W image (lower panel).
      The continuous lines are the respective best-fit models, convolved with the
      instrumental PSF and with a constant background level added.
      For the parameters of the best-fit models see text.}
         \label{modfit}
   \end{figure}

Cluster structural models were fit to the profile using the methods described
in McLaughlin et al. (\cite{mclaughlin}). Before fitting to the data, the models were
convolved with a PSF profile derived from ellipse measurements of TinyTim model
Point Spread Functions (PSFs)  for the center of the PC camera. We considered
the same three models used in Barmby et al. (\cite{bar07}): King (\cite{k66}),
Wilson (\cite{wils}), and S\'ersic (\cite{sers}). The background level  (i.e.,
the intensity of the largest isophotes) was allowed to vary in the fitting.
Fig.~\ref{modfit}  shows the profile data and the best-fit models in the two
filters. Small scale bumps in the observed profile are likely due to individual
bright stars (SGs). For the F450W filter the S\'ersic model with index $n=4.0$
was the best fit. This model has central intensity  $I_0 = 7.9\times10^5
L_{\sun}$~pc$^{-2}$ and scale radius $r_0 = 6.1\times10^{-4}$~pc. The projected
half-light radius is $r_h = 9.12~$pc (2\farcs40) and total luminosity
(corrected for extinction) $1.5\times10^6 L_{\sun}$. For the F814W image, the
best-fit model was a Wilson (\cite{wils}) model with $W_0=11.2$,  central
intensity  $I_0 = 5.0\times10^5 L_{\sun}$~pc$^{-2}$ and scale radius $r_0 =
0.072$~pc. The projected half-light radius is $r_h = 5.60~$pc (1\farcs47) and
total luminosity $5.7\times10^5 L_{\sun}$.  In the following analysis, we adopt
the average of the two half-light radii, $r_h = 7.4 \pm 2.5~$pc (1\farcs94
$\pm$ 0\farcs66; the reported uncertainty is the standard deviation of the two
values).  It is also interesting to note that the half-light radius we have
derived for
VdB0 is larger than those for the clusters listed by Figer (see Davies
et al.~\cite{davies}, $r_h\simeq 0.2 - 3$pc), but smaller than  NGC~1850
($r_h\simeq 13$ pc) and very similar to NGC~1711  ($r_h\simeq 6$ pc),  for
example\footnote{The surface brightness profiles of these and other LMC
clusters have been studied by Mackey \& Gilmore (\cite{mack}) who provide  the
parameters of the EFF87 models that  best fit the observed profiles. To derive
the reported half-light radii we searched for the King \cite{k62}  model
providing the best match to the EFF87 best-fit profile found by  Mackey \&
Gilmore (\cite{mack}), and adopted the corresponding $r_h$.}. A summary of the
adopted structural parameters of VdB0 is reported in Table~\ref{table:3}.

%
%


The derived values of the total luminosity  correspond
to $M_{450W}=-10.13$ and $M_{814W}=-10.25$, respectively.
Using Eq.~12 of Dolphin (\cite{dolcal}) these VEGAMAG magnitudes can be
transformed to standard $B$ and $I$ using the appropriate coefficients from
his Table~7. 
The integrated $(B-V)_0$ color required for the
transformation has been taken from the RBC ($(B-V)_0=0.05$, 
Tab.~\ref{table:1}, above), 
while we adopted $(V-I)_0=0.40$ from Maraston's (\cite{mara2}) model 
for a solar metallicity
Simple Stellar Population (SSP\footnote{A Simple Stellar Population is a
population of stars all having the same age and chemical composition and having
individual masses extracted from a given Initial Mass Function (IMF); this
is a practical idealized model that is generally believed to be a reasonable
approximation of a star cluster, see Renzini \& Fusi Pecci \cite{rfp}.}) 
with age of 25~Myr, as an observational estimate of the $I$ magnitude of VdB0 
was not available (but see below).
$M_V=-9.9$ is obtained from $M_{814W}$ and $M_V=-10.2$ from $M_{450W}$;
we adopt the average (in flux) of the two, $M_V=-10.06$.
This value is in excellent agreement with the value of $M_V=-10.03$ 
listed in the RBC, and coming, in turn, from the photometry by Sharov 
et al. (\cite{shar}). 

There are, however,
compelling reasons to consider the
estimate of $M_V$ obtained from our HST images as significantly uncertain
because of the unfortunate combination 
of a very extended cluster and of
a very low intrinsic background level
(just 1 to 2 DN in the background sky in the original raw
WFPC2 images, particularly for the F450W filter).
This  guarantees
that photometry within very large apertures will have a large
uncertainty, and the resulting integrated brightness 
may depend on the details of how the code 
handles the background estimate in this photon-starved regime.

For this reason we prefer to rely on the excellent ground-based material that
is publicly available to obtain a reliable estimate of the total luminosity of
the cluster.  Existing ground-based photometry 
of VdB0 taken from Sharov et al. (\cite{shar}) is compiled in the RBC.  
However, it is possible
that it 
was obtained adopting apertures that were not large enough to 
include the whole light distribution of this particularly extended cluster
(see Fig.~\ref{prof} and ~\ref{modfit}).  We have therefore used two
independent and well calibrated
publicly available imaging surveys covering M31 to determine
the integrated brightness of the cluster VdB0, that of 
Massey et al. (\cite{massey}, hereafter M06) and the Sloan
Digital Sky Survey (SDSS).  In both cases we
use an aperture with r=14\farcs4.  From the BVRI images of the former we
obtained $B=14.94\pm 0.09$, $V=14.67\pm 0.05$, $R=14.45\pm 0.11$ and 
$I=14.01\pm 0.11$\footnote{We note that these values imply
$(V-I)_0=0.41$, adopting the reddening law by 
Dean, Warren \& Cousins~(\cite{dean}), in excellent agreement with the 
prediction, used above, of $(V-I)_0$ from Maraston's (\cite{mara2}) model for 
a solar metallicity SSP of age 25 Myr.}.
The SDSS - Data Release 6 (DR6, Adelman-McCarthy et al. \cite{dr6})
$g$, $r$, and $i$ images yielded $B=14.92$, $V=14.63$, $R=14.45$, and $I=14.03$
using the color transformations of
Lupton (\cite{lupton}), in excellent agreement with those inferred
from the M06 images.   This is $\simeq 0.4 - 0.6$~mag brighter
than those reported in the RBC.
In Sect.~4 we will show that the J,H,K magnitudes of VdB0 also become
brighter by $\sim 0.2-0.5$ mag after increasing the adopted aperture from 
$r=$5\farcs0 to 15\farcs0. 

Given all the above, we adopt  the $r=$14\farcs4 aperture photometry measured on
M06 images as our preferred values, reported in Table~\ref{table:3}, below. 
In particular $V=14.67\pm0.05$ is our final best estimate of
the integrated V magnitude of VdB0, corresponding to $M_V=-10.42 \pm 0.20$;
these values will be  adopted in the following analysis. 

\section{Summary and discussion}

We have outlined the data reduction and scientific analysis strategy that we
adopt for our HST-WFPC2 survey of M31 candidate YMCs, 
whose complete results will be presented in future contributions.
As an exemplary case, we have described
the study of the cluster VdB0. We have found that VdB0 is a very bright and
extended cluster of approximately solar metallicity and of age $\sim 25$ Myr,
with a rich population of blue and red supergiants. 

   \begin{figure}
   \centering
   \includegraphics[width=9cm]{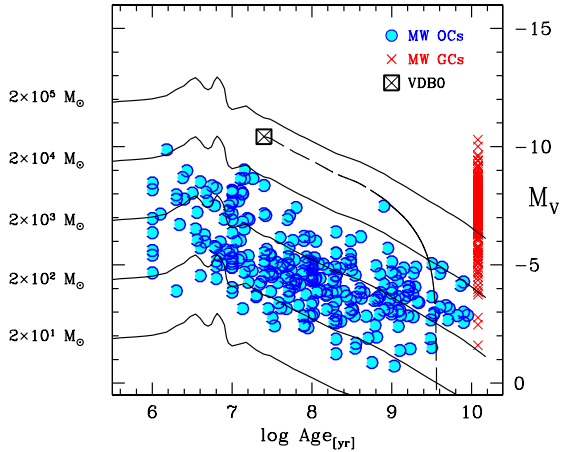}
      \caption{Integrated V mag and total mass as a function of age for 
      various clusters. Galactic Open Clusters (OC, from the 
      WEBDA database) are plotted filled circles, 
      Galactic Globular Clusters (GC, $M_V$ from the most recent version of the 
      Harris (1996) catalogue, i.e. that of February 2003; the ages have been
      arbitrarily assumed to be 12.0~Gyr for all the clusters) are plotted as
      $\times$ symbols.      
      VdB0 is represented as a crossed square at $M_V=-10.42$, from 
      Tab.~\ref{table:3}.
      The continuous lines are fixed-stellar-mass models from the set
      by Maraston (1998, 2005) for SSPs of solar metallicity, with a
      Salpeter's Initial Mass Function (IMF) and intermediate Horizontal Branch
      morphology. Note that in this plane, the dependence of the models from 
      the assumed IMF, metallicity and HB morphology is quite small 
      (see B08). 
      The outlier OC at log Age$\simeq 9.0$ is Tombaugh~1.
      The long dashed line is the VDB0 evolutionary
      track including the mass loss by dynamical effects according to
      the formulas by LG06. The cluster is
      expected to dissolve within $<4$~Gyr from the present epoch.}
         \label{vdb0}
   \end{figure}
%

Having clearly ascertained that VdB0 is a real cluster, it remains to be
established if it is more similar to ordinary open clusters of the Milky Way 
than to to the Young Massive Clusters that may be considered
as possible precursors of ``disk globulars''.
The similarity with LMC objects  typically classified
as ``Young Globular Clusters''  such as NGC1850 (see Sect.~3., above)  is
quite remarkable and it suggests that VdB0 is not an ordinary 
OC (but see also point 1, below).

A more
general way to compare clusters of different ages, taking  
into account the
fading of the luminosity of SSPs as they age, it is to plot them into a diagram
comparing age to some indicator of the stellar mass of the cluster (see, for
example,  Whitmore, Chandar \& Fall \cite{whitmore}, Gieles, Lamers \&
Portegies-Zwart \cite{gieles}, and de Grijs, Goodwin \& Kouwenhoven
\cite{degris}, for recent applications and references). 
Here we adopt 
log(Age) vs. absolute integrated magnitude as in B08.

   \begin{figure}
   \centering
   \includegraphics[width=9cm]{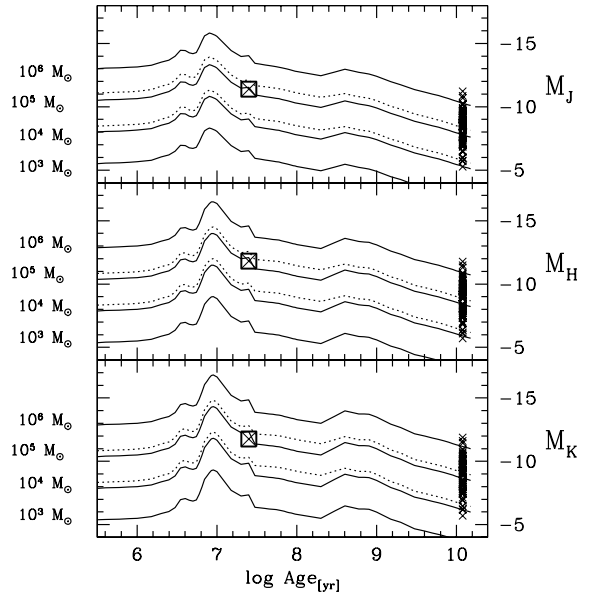}
      \caption{The same as Fig.~\ref{vdb0} but for near infrared colors.
      Integrated magnitudes of GCs are taken from Cohen et al. (\cite{cohir});
      the IR magnitudes for VdB0 are taken from Tab.~4.
      The dotted lines are $M=10^4 M_{\sun}$ and $M=10^5 M_{\sun}$
      iso-mass models assuming a Kroupa \cite{kroupa} IMF instead of a 
      Salpeter (\cite{salp}) IMF, plotted here to illustrate the weak effect 
      of assumptions on IMFs.}
         \label{mir}
   \end{figure}
%

In Fig.~\ref{vdb0} VdB0 is compared with Galactic Open Clusters (data taken
from the WEBDA 
database\footnote{\tt http://www.univie.ac.at/webda/integre.html}),
with  Galactic
Globular Clusters (from the latest version of Harris \cite{h96}  assuming
a uniform age of 12 Gyr, a reasonable
approximation for our purpose), and with a
grid of SSP models with solar metallicity and Salpeter's IMF from the set by
Maraston\footnote{\tt http://www-astro.physics.ox.ac.uk/~maraston/} 
(\cite{mara1,mara2}). 
As a SSP ages massive stars die while the mass of the most luminous stars 
decreases (passive evolution). 
Keeping the total mass fixed, the luminosity
of the population fades and, as a consequence, the stellar mass-to-light (M/L) 
ratio increases. The continuous lines plotted in Fig.~\ref{vdb0} describe
the passive evolution of SSPs of various (stellar) masses: under the adopted
assumptions the mass of a cluster of given age and $M_V$ can be read from
the grid of iso-mass tracks. 

The path of the track passing through the cluster
shows what  its luminosity will be in the future if the cluster did not
lose stars through dynamical processes (evaporation, tides, ecc.). 
The latter is clearly not the case in general, and in particular for VdB0. 
In addition to the relatively mild
evaporation driven by two body encounters, 
it will suffer from the strain of the
M31 tidal field and from  encounters with 
Giant Molecular Clouds (GMC), 
as the cluster is embedded in the dense thin disk of M31 
(Lamers \& Gieles \cite{lamers}, hereafter LG06, and references therein). 
To take these effects into account we used the analytical approach presented 
by LG06 to produce an evolutionary track including the cluster mass loss by
stellar evolution, galactic tidal field, spiral arm shocking, and encounters
with giant molecular clouds, plotted in Fig.~\ref{vdb0} as a 
long-dashed curve. The LG06 formulas describe the evolution of a cluster 
located within the Milky Way (thin) disk at the Solar circle. 
They should provide a reasonable approximation for VdB0 which lies in the disk
of M31,
at a similar distance from the center of a similarly massive spiral galaxy 
(van den Bergh \cite{vdbbook}). The required inputs are the cluster mass, for
which we adopted the value that can be read from the SSP grid of
Fig.~\ref{vdb0} (see below), and the half-light radius, which we obtained in
Sect.~3.3, above (see Tab.~\ref{table:3}). The initial expulsion of gas not
used in star formation may lead young clusters (age $<50$ Myr) to lose their
virial equilibrium and it may represent an additional relevant factor driving 
toward the
destruction of clusters like VdB0 that is not included in the LG06
approach (Bastian \& Goodwin \cite{bastian6};
Goodwin \& Bastian \cite{goodwin}; Bastian et al. \cite{bastian8}).

\begin{figure}
   \centering
   \includegraphics[width=9cm]{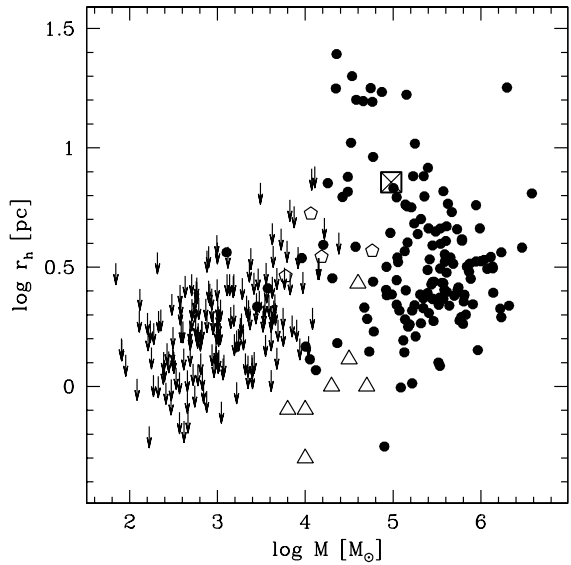}
      \caption{VdB0 (crossed square) is compared to other clusters in 
      the {\em logarithm of
      the mass} vs. {\em logarithm of the half-light-radius} plane.
      Filled circles are Galactic GCs from Mackey
      \& van den Bergh (\cite{macksyd}). 
      Arrows are Galactic OCs: we plot the
      radii where a break in the surface brightness profile occurs, taken
      from Kharchenko et al. (\cite{kar}, their ``core radii'').
      These should be considered as upper
      limits for actual $r_h$, which are not available for most OCs.
      The masses of the OCs have been computed using the grid of SSP models
      shown in Fig.~\ref{vdb0}, while for GCs we adopted age=12.0 Gyr and
      a grid of SSP models having $[Z/H]=-1.35$. 
      Open pentagons are the clusters studied by WH01.
      Open triangles are the massive young MW clusters listed by 
      Figer (\cite{figer}); masses and radii are taken
      from his Table~1. Note that the radii reported by Figer for these 
      clusters are not half-light-radii, however they should be a reasonable
      proxy. 
      The good match between the two quantities has been verified in 
      the case of RSG1, for which Figer report $r = 1.3$~pc, and Davies et 
      al. (\cite{davies}) obtain $r_h=1.5\pm 0.3$ pc.}
         \label{mvrhN}
   \end{figure}

Fig.~\ref{vdb0} is worth of some detailed considerations:

\begin{enumerate}

\item{} Independently of the exact value of $M_V$ adopted, 
 VdB0 is significantly brighter ($\ga 1$ mag) than Galactic OCs of similar ages, 
actually it is brighter than
Galactic OCs of {\em any} age. The same is true also if 
all other known M31 OCs are considered
(Hodge \cite{open}; Krienke \& Hodge \cite{krie1,krie2}).
However it should be noted that the population of disk clusters in M31 may
be so huge ($\sim 80000$ clusters, according to Krienke \& Hodge \cite{krie1})
that even the extreme tails of the luminosity distribution may be populated. (This
should not be the case for the LMC, for example, as it is orders of
magnitude less massive than M31). Hence it is premature to draw a conclusion
from an individual cluster; when the whole sample is analyzed we will get 
a deeper insight on the actual nature of VdB0. 

\item{} Assuming the RBC value for the integrated V magnitude, E(B-V)=0.0
instead of E(B-V)=0.2 and a grid of iso-mass tracks adopting a Kroupa IMF, 
we can obtain an extremely conservative strong {\em lower limit} to the 
stellar mass 
of VdB0, $M=2.4\times 10^4 M_{\sun}$. Under the same assumptions but 
adopting the best-fit value  E(B-V)=0.2 we
obtain $M=6.5\times 10^4 M_{\sun}$ with a Salpeter IMF and
$M=4.2\times 10^4 M_{\sun}$ with a Kroupa IMF. These are at the threshold
between the OC and GC mass distributions (see van den Bergh \&
Lafontaine \cite{vdblf} and B08) and also at the
upper end of the mass distribution of Galactic YMC (see Figer \cite{figer} and
Fig.~\ref{mvrhN}, below).
The conclusion that VdB0 is much more massive than MW clusters
of similar ages seems inescapable, unless extreme IMFs are considered 
(i.e. IMF truncated at low masses, see Sternberg \cite{stern}).

\item{} If $M_V=-10.42$ is adopted, as obtained from large aperture ground-based
V photometry in Sect.~3.4, the total stellar 
mass is  $M=9.5\times 10^4 M_{\sun}$ with a Salpeter IMF and
$M=6.0\times 10^4 M_{\sun}$ with a Kroupa IMF. 

\item{} The evolutionary tracks including the LG06 treatment of mass-loss by
dynamical effects show that, independent of the actual mass 
(within the range outlined above), it is unlikely that the cluster VdB0 
would survive  for an Hubble time.  Hence 
it is very probable that it will never have the opportunity to evolve into a
classical (faint) GC. The disruption timescale is dominated by encounters
with GMCs; considering this effect alone (Eq.~7 of LG06) the cluster is
predicted to dissolve within $\simeq 3.6$ Gyr if its mass is  
$M=9.5\times 10^4 M_{\sun}$, as obtained from our best estimate of the
integrated V magnitude and assuming a Salpeter's IMF. 

\item {} In the same grid of Fig.~\ref{vdb0} and under the same
assumptions the masses of the BLCCs observed by WH01 - adopting their age
estimates -  range  from $8.0\times 10^3 M_{\sun}$, (G293) in the realm of OCs,
to $\simeq 2 \times 10^4 M_{\sun}$ (G44 and G94) and 
$8\times 10^4 M_{\sun}$ (G38), very similar to that of  VdB0
and significantly larger than OCs of similar ages. 

\end{enumerate}

To obtain independent and more robust estimates of the present-day stellar mass
of VdB0 we used the Near Infrared (NIR) version of the log~Age vs. absolute
integrated magnitude plane. In Fig.~\ref{mir}, J,H and K absolute magnitudes of
VdB0 extracted from the Extended Sources Catalogue (XSC) of 2MASS
are compared with Maraston's SSP models of solar metallicity and Salpeter's 
(continuous lines) or Kroupa's (dotted lines) IMFs and with Galactic GCs 
(from Cohen et al. \cite{cohir}, ages assumed as above)\footnote{For J,H,K colors 
we adopt  $A_J=0.871E(B-V)$, $A_H=0.540E(B-V)$, and $A_K=0.346E(B-V)$, from
Rieke \& Lebofsky \cite{rieke}. The J,H,K absolute magnitudes of the Sun are 
taken from Holmberg, Flynn \& Portinari \cite{holm}.}. 
NIR integrated magnitudes for significant samples of OCs are not available, at 
present. 
To account for the whole extent of the cluster we extracted $r=15\arcsec$ 
aperture photometry, that is provided in the XSC, instead of the $r=5\arcsec$ 
adopted in the RBC, see Tab.~\ref{table:1} and Tab.~\ref{table:3}).

NIR magnitudes are more reliable mass tracers than
visual magnitudes as NIR M/L ratios are smaller and have smaller variations
with age, compared to optical M/L ratios. For example, according to  Maraston
(\cite{mara1,mara2}) models, a solar metallicity Salpeter-IMF SSP at Age = 10
Gyr has (M/L)$_V$=5.5, while  (M/L)$_K$=1.4; the same SSP has
$\frac{d(M/L)_V}{dt}\simeq 0.55$ while  $\frac{d(M/L)_K}{dt}\simeq 0.13$.
The independent estimates of the stellar mass from J,H, and K magnitudes 
are essentially identical, ranging
from 6 to 9 $\times 10^4 M_{\sun}$, assuming a Salpeter IMF,  and  from 4 to
5.5 $\times 10^4 M_{\sun}$, assuming a Kroupa IMF.  These estimates are in 
fair agreement with those obtained from the integrated V photometry.

%

%
\begin{table}
\caption{Newly derived coordinates, half-light radius, integrated magnitudes,
reddening, age and metallicity for the cluster VdB0. The origin of each
parameter is described in the last column.}             
\label{table:3}      
\centering                          
\begin{tabular}{c c c}        
\hline\hline                 
par & value & note \\    
\hline   
&&\\
$\alpha_{J2000}$& $00^h$ $40^m$ $29.4^s$               & from 2MASS-XSC\\                    
$\delta_{J2000}$& $+40\degr$ $36\arcmin$ $15.2\arcsec$ & from 2MASS-XSC \\                  
\hline                        
&&\\
$r_h$       & 1\farcs93 $\pm$ 0\farcs66  & from intensity profile (i.p.)
fit \\    
\hline
&&\\
B     &  14.94$\pm$ 0.09  & r=14\farcs4 ap. phot. on M06 images\\                               
V     &  14.67$\pm$ 0.05  & r=14\farcs4 ap. phot. on M06 images\\                               
R     &  14.45$\pm$ 0.11  & r=14\farcs4 ap. phot. on M06 images\\                               
I     &  14.01$\pm$ 0.11  & r=14\farcs4 ap. phot. on M06 images\\                               
J     &  13.26$\pm$ 0.07  & r=15\farcs0 ap. phot. from 2MASS-XSC\\                               
H     &  12.76$\pm$ 0.12  & r=15\farcs0 ap. phot. from 2MASS-XSC\\                               
K     &  12.77$\pm$ 0.15  & r=15\farcs0 ap. phot. from 2MASS-XSC\\                               
\hline    
\hline    
&&\\
age & 25 Myr               & value of adopted best-fit isochrone\\ 
Z   & 0.019                & value of adopted best-fit isochrone\\                             
E(B-V)   & 0.20            & adopted best-fit value\\                             
\hline    
\end{tabular}
\end{table}

Finally, in Fig.~\ref{mvrhN} we compare VdB0 with Galactic OCs, GCs and YMC,
plus the BLCCs studied by WH01, in the log of the {\em stellar mass} versus 
log of the half-light radius plane (similar to Mackey \& 
van den Bergh \cite{macksyd} and  Federici et al. \cite{luckyrt}). 
The radii at which the break in the
profile (core/corona transition) of Galactic OCs (from Kharchenko et al.
\cite{kar}) occurs is taken as a strong upper limit for their $r_h$. 
VdB0 has a typical size that is larger than both OCs and YMCs, 
and is similar to that of several MW GCs of comparable mass.

In conclusion, we can say that VdB0 seems a remarkable cluster
 in several of its properties when compared
to the other known disk clusters of the Milky Way and M31. In this paper we
have presented the data reduction, data analysis and diagnostics that will be
applied to the whole survey sample and that will allow us to put VdB0 and the
other clusters in the more general context of the star cluster populations in
the disk of spiral galaxies.  

\begin{acknowledgements}
S.P. and M.B. acknowledge the financial support of INAF through the 
PRIN 2007 grant CRA 1.06.10.04 ``The local route to galaxy
formation...''. 
P.B. acknowledges research support through a Discovery Grant from the Natural 
Sciences and Engineering Research Council of Canada.
J.G.C. is grateful for partial support through
grant HST-GO-10818.01-A from the STcI.
THP gratefully acknowledges support in  
form of a Plaskett Fellowship at the Herzberg institute of  
Astrophysics in Victoria, BC.
J.S. was supported by  NASA through an Hubble Fellowship, 
administered by STScI. We are grateful to S. van den Bergh for having pointed 
out some errors in the historical reconstruction of the discovery of VdB0 that 
were  reported in a previous version of the paper.
\end{acknowledgements}

\end{document}